\documentclass[epsf,amssymb,aps,prd,nofootinbib,floatfix,superscriptaddress,secnumarabic]{revtex4}

\usepackage{bm}
\usepackage{amsmath}                                                      
\usepackage{graphicx}
\usepackage{epsf}
\usepackage{epsfig}
\usepackage{epstopdf}
\usepackage{hyperref}
\usepackage{amsmath}
\usepackage{amsfonts,amssymb}
\usepackage{colordvi}
\usepackage{color}
\usepackage{lscape}
\usepackage{rotfloat}
\usepackage{rotating}
\usepackage{slashed}
\usepackage{float}
\usepackage{array}
\usepackage{dsfont}
\usepackage{booktabs}
\usepackage[english]{babel}

\newcommand{\be}{\begin{equation}}
\newcommand{\ee}{\end{equation}}
\newcommand{\bea}{\begin{eqnarray}}
\newcommand{\eea}{\end{eqnarray}}

\newcommand{\MSbar}{{\overline{\rm MS}}}

\newcommand{\Tr}{{\rm Tr}}

\newcolumntype{M}[1]{>{\centering\arraybackslash}m{#1}}
\newcolumntype{N}{@{}m{0pt}@{}}

\def\lsim{\mathrel{\rlap{\lower4pt\hbox{\hskip1pt$\sim$}}
    \raise1pt\hbox{$<$}}}                
\def\slashed{{/}\mskip-10.0mu}

\begin{document}
\vspace*{1cm}

\title{Gauge-invariant Renormalization Scheme in QCD: Application to fermion bilinears and the energy-momentum tensor}

\author{M.~Costa}
\email{kosta.marios@ucy.ac.cy}
\affiliation{Department of Physics, University of Cyprus, Nicosia, CY-1678, Cyprus}
\affiliation{Department of Mechanical Engineering and Materials Science and Engineering, Cyprus University of Technology, Limassol, CY-3036, Cyprus}

\author{I.~Karpasitis}
\email{i.karpasitis@cyi.ac.cy}
\affiliation{Department of Physics, University of Cyprus, Nicosia, CY-1678, Cyprus}
\affiliation{Present address: The Cyprus Institute, Nicosia, CY-2121, Cyprus}

\author{G.~Panagopoulos}
\email{gpanago@stanford.edu}
\affiliation{Department of Physics, Stanford University, California, 94305–2004, USA}

\author{H.~Panagopoulos}
\email{haris@ucy.ac.cy}
\affiliation{Department of Physics, University of Cyprus, Nicosia, CY-1678, Cyprus}

\author{T.~Pafitis} 
\email{pafitis.theodosis@ucy.ac.cy}
\affiliation{Department of Physics, University of Cyprus, Nicosia, CY-1678, Cyprus}
\affiliation{Present address: Department of Physics, Utrecht University, 3508 TC Utrecht, the Netherlands}

\author{A.~Skouroupathis}
\email{askour02@ucy.ac.cy}
\affiliation{Department of Physics, University of Cyprus, Nicosia, CY-1678, Cyprus}
\affiliation{Cyprus Ministry of Education, Culture, Sport and Youth, Nicosia, CY-1434, Cyprus}

\author{G.~Spanoudes}
\email{spanoudes.gregoris@ucy.ac.cy}
\affiliation{Department of Physics, University of Cyprus, Nicosia, CY-1678, Cyprus}

\begin{abstract}  
We consider a gauge-invariant, mass-independent prescription for
renormalizing composite 
operators, regularized on the lattice, in the spirit of the coordinate
space (X-space) renormalization scheme. The prescription involves only
Green's functions of products of gauge-invariant operators, situated
at distinct space-time points, in a way as to avoid potential contact
singularities. Such Green's functions can be computed
nonperturbatively in numerical simulations, with no need to fix a
gauge: thus, renormalization to this ``intermediate'' scheme can be carried
out in a completely nonperturbative manner.

Expressing renormalized operators in the $\MSbar$ scheme
requires the calculation of corresponding conversion factors. The
latter can only be computed in perturbation theory, by the
very nature of $\MSbar$; however, the computations
are greatly simplified by virtue of the following attributes:

i) In the absense of operator mixing, they involve only massless, two-point
functions; such quantities are calculable to very high perturbative order. 

ii) They are gauge invariant; thus, they may be computed in a
convenient gauge (or in a general gauge, to verify that the result is
gauge-independent).

iii) Where operator mixing may occur, only gauge-invariant operators
will appear in the mixing pattern: Unlike other
schemes, involving mixing with gauge-variant operators (which
may contain ghost fields), the mixing
matrices in the present scheme are greatly reduced. 
Still, computation of some three-point functions may not be altogether
avoidable.

We exemplify the procedure by computing, to lowest order, the
conversion factors for fermion bilinear operators of the form
$\bar\psi\Gamma\psi$ in QCD.

We also employ the gauge-invariant scheme in the study of 
mixing between gluon and quark energy-momentum tensor operators: 
We compute to one loop the conversion factors relating the nonperturbative 
mixing matrix to the $\MSbar$ scheme.

\end{abstract}

\hspace{-5mm}\begin{minipage}{\textwidth}
\maketitle
\end{minipage}
\renewcommand{\thefootnote}{\alph{footnote}}
\footnotetext{Electronic addresses: ${}^{a}$ kosta.marios@ucy.ac.cy, \ ${}^{b}$ i.karpasitis@cyi.ac.cy , \ ${}^{c}$ gpanago@stanford.edu, \ ${}^{d}$ haris@ucy.ac.cy, \\
${}^{e}$ pafitis.theodosis@ucy.ac.cy, \ ${}^{f}$ askour02@ucy.ac.cy, \ ${}^{g}$ spanoudes.gregoris@ucy.ac.cy}
\renewcommand{\thefootnote}{\arabic{footnote}}

\section{Introduction}
\label{Introduction}
\bigskip

Renormalization of composite operators is essential when studying matrix elements and correlation functions in Hadronic Physics. It relates bare quantities of the theory to the physical ones. In order to extract nonperturbative physical results from numerical simulations on the lattice, the construction of a proper nonperturbative renormalization scheme is needed. A requirement for such scheme is to be applicable in both continuum and lattice regularizations in order to make contact with the continuum schemes. Nowadays, the most widely used renormalization scheme in lattice simulations is the modified regularization-independent scheme (RI$'$)~\cite{Bochicchio:1985xa,Martinelli:1994ty}; it considers gauge-variant Green's functions (GFs) of composite operators with external elementary quantum fields in momentum space. This is not a unique nonperturbative scheme. In this paper, we consider an alternative approach, which involves gauge-invariant correlation functions of composite operators in coordinate space. This approach, called ``X-space'' scheme, has been considered in~\cite{Gimenez:2004me} in the context of lattice studies some years ago. Older investigations of coordinate-space methods can be also found in, e.g.,~\cite{Jansen:1995ck}. To date, there are only limited lattice applications of the X-space scheme, mainly regarding the multiplicative renormalization of fermion bilinear operators. Other applications regarding more complex operators, such as the four-fermion operators, have been studied in, e.g.,~\cite{Dimopoulos:2018zef}. X-space is a promising nonperturbative scheme for the lattice simulations, especially when one considers further applications involving operators which mix under renormalization. However, some extensions are necessary in order to deal properly with the error in nonperturbative calculations and, most importantly, with operator mixing. In this paper we implement a number of extensions to this effect; the resulting scheme will be referred to as ``Gauge-Invariant Renormalization Scheme (GIRS)'' to emphasize the property of gauge invariance, which is essential when one studies the renormalization of gauge-invariant operators in the presence of mixing.

GIRS involves two-point GFs of the following form: \\
\begin{equation}
\langle \mathcal{O}_1 (x) \mathcal{O}_2 (y) \rangle, \qquad (x \neq y),
\end{equation}
where $\mathcal{O}_1 (x), \mathcal{O}_2 (y)$ are gauge-invariant operators at two different spacetime points. In many cases the renormalization factors of the operators in GIRS can be extracted by studying only two-point functions; however, as we conclude by this work, in the presence of mixing the study of three-point functions is also needed in numerous cases. This scheme has a number of advantages which make easier its implementation in the lattice simulations:
\begin{enumerate}
\item The GFs under consideration are gauge-invariant. The benefits from this property are: Firstly, when mixing occurs, the set of operators under mixing is reduced. Gauge-variant operators [BRST (Becchi-Rouet-Stora-Tyutin) variations and operators which vanish by the equations of motion] which mix with gauge-invariant operators (according to the Joglekar-Lee theorems~\cite{Joglekar:1975nu}) do not contribute in these GFs. This property is very useful especially when studying the renormalization of gauge-invariant operators nonperturbatively by lattice simulations; gauge-variant operators, typically, contain ghost fields and/or gauge-fixing terms, which are defined in perturbation theory and their study is not obvious in a nonperturbative context. Secondly, no gauge fixing is needed in GIRS. When fixing a covariant gauge on the lattice, one encounters the problem of Gribov copies (see, e.g.,~\cite{Maas:2011se, Cucchieri:2018doy}). Employing this scheme, we avoid such a problem. Given that GFs in GIRS are independent of the gauge-fixing parameter, one can perform perturbative calculations in the Feynman gauge, where momentum-loop integrals are simpler.
\item Contact terms are automatically excluded ($x \neq y$) in contrast to the standard renormalization schemes in momentum space (e.g., RI$'$ scheme). 
\item In the absence of operator mixing, perturbative calculations in GIRS involve diagrams with only one inconing/outgoing momentum. Given that one may also adopt a massless renormalization scheme, one can make use of techniques for evaluating such diagrams which have been developed to very high perturbative order (see, e.g.,~\cite{Chetyrkin:1999pq, Ruijl:2017eht, Luthe:2017ttg, Chetyrkin:2017bjc, 2018PhRvD..97h5016G, 2018arXiv181211818H, Baikov:2019zmy}). Analogous techniques can be used even in the presence of mixing.
\item The fact that GIRS renormalization functions can be fully obtained non-perturbatively, without recurrence to lattice perturbation theory, has the consequence that conversion to the modified minimal subtraction ($\MSbar$) scheme entails only continuum perturbative calculations.
\end{enumerate}

There are also some disadvantages of GIRS:
\begin{enumerate}
\item Computations in GIRS, at a given order in perturbation theory, involve diagrams with one more loop. 
\item The cost for not generating contact terms is the presence of exponentials in Feynman integrals, which makes their computation somewhat more complex.
\item When mixing occurs, one must often study also ($n>2$)-point GFs; this is of course the case not only in GIRS, but also in other schemes.
\end{enumerate}

Both RI$'$ and GIRS are used as intermediate schemes in which renormalization functions can be directly obtained via lattice simulations. The ultimate goal is to obtain renormalized GFs in the $\MSbar$ scheme, which is the standard scheme used in the analysis of experimental data. For this purpose, one must compute appropriate conversion factors, which are finite and regularization independent.

Our work is divided into two parts: The first part focuses on the employment of GIRS in the multiplicative renormalization of fermion bilinear operators. This serves both as an example for describing the renormalization procedure in GIRS and as a necessary ingredient in possible variants of GIRS appearing in the second part. The fermion bilinear operators have already been studied in GIRS in both continuum~\cite{Gracey:2009da,Chetyrkin:2010dx} and lattice regularizations~\cite{Gimenez:2004me,Cichy:2012is,Tomii:2018zix}. A novel aspect of our calculation is that we provide alternative ways of implementing GIRS: i.e. using specific values of $x, y$, or integrating over timeslices. Such choices might help to reduce statistical noise in the nonperturbative evaluation. Given that the $\MSbar$ renormalization functions are independent of the spacetime points $x, y$, the nonperturbative estimates can be checked by verifying this property. Another aspect is that we also provide results using the t'Hooft-Veltman d-dimensional definition of $\gamma_5$. 

In the second part of our work, we extend the application of GIRS in the presence of mixing: we study the renormalization and mixing of the gluon and quark parts of the QCD energy-momentum tensor (EMT); this is a subject of research with an increased interest in recent years~\cite{Yang:2018bft, Shanahan:2018pib, Yang:2018nqn, Alexandrou:2020sml, DallaBrida:2020gux}. EMT is relevant to the calculation of the renormalized gluon and quark average momentum fractions, which are involved in the study of hadron spin decomposition~\cite{Ji:1998pc}. In this study, we consider only nondiagonal elements of EMT, which give a simpler mixing pattern. However, the procedure can be similarly extended to the renormalization of the diagonal elements. In order to establish the required number of renormalization conditions we must also consider three-point GFs. A number of candidate GFs can be employed for this purpose, such as: $\langle {\cal O}_i^{\mu \nu} (x) {\cal O}_{\openone} (y) {\cal O}_{\openone} (z) \rangle$, where ${\cal O}_i^{\mu \nu}$ is the gluon or quark energy-momentum tensor operator and ${\cal O}_{\openone} (x) = \bar{\psi} (x) \psi (x)$ is the scalar bilinear operator; we compute the corresponding conversion factors for some of the most prominent candidates.       

The outline of this paper is as follows: Sec.~\ref{Renormalization of fermion bilinear operators in GIRS} regards the renormalization of fermion bilinear operators in GIRS, while Sec.~\ref{Renormalization and mixing of the quark and gluon energy-momentum tensor operators in GIRS} is devoted to the renormalization and mixing of the quark and gluon EMT operators. In both sections, we provide details on the calculational procedure and we present our tree-level and one-loop results for the bare GFs of operators under study, as well as for the conversion factors between GIRS and $\MSbar$ schemes. Finally, we conclude in Sec.~\ref{summary} with a summary of our calculation and possible future extensions of our work. We also include an appendix containing details on technical aspects of the calculation (Appendix~\ref{Technical aspects of the calculation}).

\section{Renormalization of fermion bilinear operators in GIRS}
\label{Renormalization of fermion bilinear operators in GIRS}

\subsection{Details of the Calculation}
\label{Details of the Calculation}

Definitions of renormalization factors are related to specific renormalization schemes. From the perturbative point of view, these factors depend on properties of the scheme, such as the renormalization scale, the regulator, and the imposed renormalization conditions on GFs. These conditions connect bare and renormalized quantities. There exists a variety of schemes for defining renormalized quantities. A gauge-invariant renormalization scheme, which is not strictly perturbative as is $\MSbar$, is GIRS.  In order to determine renormalization factors within GIRS, we will examine GFs which contain the product of two gauge-invariant composite operators, defined at different spacetime points in the massless limit. Such GFs, computed in a lattice simulation, will lead to a nonperturbative determination of the renormalized composite operators in this scheme. Having two different schemes for the same regulator, we can calculate conversion factors between these two schemes. The renormalized quantities in one scheme can be obtained as functions of their values in the other scheme, with the bare quantities being the same in both schemes. Perturbatively, we use the dimensional regularization (DR) and we calculate these conversion factors, which can take us from GIRS to the $\MSbar$ scheme. Although we will be presenting results only up to first order beyond the leading contribution, the fact that these conversion factors can be calculated in DR, reinforces the prospect of evaluating higher-order contributions. To this end, we calculate, in QCD, the following GF for the case of two local fermion bilinear operators ${\cal O}_{X}$, ${\cal O}_{Y}$:
\be
G (x-y) \equiv \langle  {\cal O}_{X}(x) {\cal O}_{Y}(y) \rangle, \qquad x \neq y,
\label{ExpValue}
\ee
where ${\cal O}_{\Gamma}(x) \equiv \bar \psi(x) \Gamma \psi(x)$, and $\Gamma = X, Y$ denotes products of Dirac matrices given in Eqs. (\ref{scalar} - \ref{tensor}). $X$ can, in principle, differ from $Y$. Note that in order to obtain a nonzero result the flavor of the fermion (antifermion) field in ${\cal O}_X$ must coincide with the flavor of antifermion (fermion) field in ${\cal O}_Y$. Depending on the choice of $\Gamma$, the operators behave under Lorentz transformations and under parity as:
\bea
\label{scalar}
 &{\cal O}_{\openone}=\bar{\psi}\psi       &  \rm{scalar},\\
\label{pseudoscalar}
 &{\cal O}_{\gamma_5}=\bar{\psi}\gamma_5\psi      & \text{pseudoscalar},\\
 \label{vector}
 &{\cal O}_{\gamma_{\mu}}=\bar{\psi}\gamma_{\mu}\psi      &  \rm{vector},\\
\label{axial_vector}
 &{\cal O}_{\gamma_5\gamma_{\mu}}=\bar{\psi}\gamma_5\gamma_{\mu}\psi     & \rm{axial \ vector},\\
\label{tensor}
 &{\cal O}_{\sigma_{\mu\nu}}=\bar{\psi}\sigma_{\mu\nu}\psi & \rm{tensor},
\eea
where $\sigma_{\mu \nu} \equiv [\gamma_\mu, \gamma_\nu] / 2$. The above composite operators, appear frequently in the study of the eigenstates of the spectrum of a theory, i.e. hadrons (see, e.g.,~\cite{Constantinou:2014tga}), and therefore it is essential to impose an appropriate renormalization scheme for them. The GF of these operators diverges as the fields are brought near each other. The choice of Eq.(\ref{ExpValue}) ensures that both the GF and the renormalized operators are independent of the gauge. On the contrary, the GFs $\langle\psi(q)\sum_{x}\mathcal{O}_{\Gamma}(x)\bar{\psi}(q') \rangle$, which are typically used for defining the RI$'$ scheme, are gauge-dependent, as they involve fundamental fields, which are not gauge-invariant.

\begin{figure}[h]
\centering
\epsfig{file=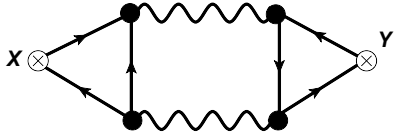,scale=0.45}
\caption{A two-loop order Feynman diagram contributing to the expectation value $\langle {\cal O}_{X}(x) {\cal O}_{Y}(y)\rangle$ for flavor singlet operators $\mathcal{O}_X$ and $\mathcal{O}_Y$. A wavy (solid) line represents gluons (quarks).}
\label{twoloop}
\end{figure}

One may consider both flavor singlet ($\frac{1}{N_f}\sum_f \bar \psi_f \Gamma \psi_f$) and nonsinglet operators ($\bar \psi_f \Gamma \psi_{f'}$, $f \neq f'$). Actually, the one-loop results do not differ between the two cases and thus we have omitted flavor indices on $\psi$, $\bar \psi$. Higher-loop contributions are expected to be different as shown in the diagram of Fig.~\ref{twoloop}.
Note that if the operators in Eq. \eqref{ExpValue} are both scalar flavor singlet, they develop a finite vacuum expectation value, which gives a mixing coefficient with the unit operator. To avoid such issues, we use normal ordered operators, i.e., ${\cal{O}}_\Gamma - \langle {\cal{O}}_\Gamma \rangle \openone$.

In our work, we extract the renormalization factors of ${\cal O}_\Gamma$, up to one loop, in both GIRS and $\MSbar$ schemes. We define the renormalized operators and parameters of the theory using the following convention:
\be
{\cal O}_\Gamma^{R} \equiv Z_\Gamma^{B,R} {\cal O}_\Gamma^{B}, \qquad g_R \equiv \mu^{(d - 4)/2} {\left(Z_g^{B,R}\right)}^{-1} g_B,
\label{RenDef}
\ee
where $g$ is the coupling constant, and $\mu$ is a momentum scale. The superscript $B$ denotes bare quantities in the $B$ regularization (e.g., $B =$ DR, LR, where DR (LR) denotes dimensional (lattice) regularization) and the superscript $R$ denotes renormalized quantities in the $R$ renormalization scheme (e.g., $R =$ GIRS, $\MSbar$). The $\MSbar$ renormalization scale $\bar{\mu}$ is defined in terms of $\mu$:
\begin{equation}
\bar{\mu} \equiv \mu {\left(\frac{4 \pi}{e^{\gamma_E}}\right)}^{1/2},
\end{equation} 
where $\gamma_E$ is Euler's gamma.

There exist several prescriptions~\cite{Larin:1993tq} for defining $\gamma_5$ in d dimensions, such as the na\"ive dimensional regularization (NDR)~\cite{Chanowitz:1979zu}, the t'Hooft-Veltman (HV)~\cite{tHooft:1972tcz}, the $DRED$~\cite{Siegel:1979wq} and the $DR{\overline{EZ}}$ prescriptions (see, e.g., Ref.~\cite{Patel:1992vu}). They are related among themselves via finite conversion factors~\cite{Buras:1989xd}. In our calculation, we apply the NDR and HV prescriptions. The latter does not violate Ward identities involving pseudoscalar and axial-vector operators in $d \equiv 4-2\,\varepsilon$ dimensions. The metric tensor, $\eta_{\mu\nu}$, and the Dirac matrices, $\gamma_\mu$, satisfy the following relations in $d$ dimensions:
\be
\eta^{\mu\nu}\eta_{\mu\nu}=d,\qquad \{\gamma_\mu,\gamma_\nu\} = 2 \eta_{\mu\nu} \openone.
\ee
In NDR, the definition of $\gamma_5$ satisfies:
\be
\{\gamma_5,\gamma_{\mu}\} = 0, \, \,\forall \mu,
\ee
whereas in HV it satisfies:
\be
\{\gamma_5,\gamma_{\mu}\} = 0, \, \,\mu = 1,2,3,4, \qquad [\gamma_5,\gamma_{\mu}]=0, \,\, \mu>4.
\ee

The renormalization factors in GIRS can be obtained by imposing the following condition: 
\begin{equation}
\langle {\cal O}^{\rm GIRS}_X (x) {{\cal O}^{\rm GIRS}_Y} (y) \rangle \vert_{x-y = \bar{z}} \ \equiv \  
Z_X^{B,{\rm GIRS}} Z_Y^{B,{\rm GIRS}} \langle {\cal O}^B_X (x) {{\cal O}^B_Y} (y) \rangle \vert_{x-y = \bar{z}} \ = \ \langle {\cal O}^{\rm GIRS}_X (x) {{\cal O}^{\rm GIRS}_Y} (y) \rangle^{\rm tree} \vert_{x-y = \bar{z}}, \label{GIRScond}  
\end{equation}
where $\bar{z}$ is the GIRS 4-vector position scale ($\bar{z} \neq (0,0,0,0)$). As we are interested in applying GIRS in lattice simulations, the scale $\bar{z}$ may be chosen to satisfy the condition $a \ll |\bar{z}| \ll \Lambda_{\rm QCD}^{-1}$, where $a$ is the lattice spacing and $\Lambda_{\rm QCD}$ is the QCD physical scale; this condition guarantees that discretization effects will be under control and simultaneously we will be able to make contact with (continuum) perturbation theory. 

There are additional, alternative ways for extracting renormalization factors in GIRS, using variants of the GFs of Eq. (\ref{ExpValue}). An option is to take a Fourier transform of Eq.(\ref{ExpValue}); however, this is not an optimal choice as contact terms arise. A more promising option is to integrate Eq.(\ref{ExpValue}) over three of the four components of the position vector $(x - y)$, while setting the fourth component equal to a reference scale $\bar{t}$. For the scalar and pseudoscalar operators, the direction of the unintegrated component is immaterial; for the other operators, there are two possible options, depending on whether this direction coincides or not with one of the indices carried by the operators. Due to the anisotropic lattice employed in simulations, the temporal direction is a special one. In this sense, a natural choice for the component $\bar{t}$ is to be temporal; we call this variant t-GIRS. Without loss of generality, we set $x=(x_1,x_2,x_3,0)$ and $y=(0,0,0,\bar{t})$; then the renormalization condition for t-GIRS takes the following form:
\begin{equation}
\int d^3\vec{x} \ \langle O^{\rm t-GIRS}_X (\vec{x}, 0) {O^{\rm t-GIRS}_Y} (\vec{0},\bar{t}) \rangle \ = \int d^3\vec{x} \ \langle O^{\rm t-GIRS}_X (\vec{x}, 0) {O^{\rm t-GIRS}_Y} (\vec{0}, \bar{t}) \rangle^{\rm tree}. \label{RC2}
\end{equation}  
Although the choice of prescription is not unique, we benefit from the fact that each choice depends on only one reference scale. In case the renormalization prescription involves GFs which are not integrated over any spacetime direction, as in Eq. \eqref{GIRScond}, the quantity $(x-y)$ can be chosen to take any 4-vector reference value, e.g. the ``democratic'' choice $\bar{z} = a (n_1, n_2, n_3, n_4)$, where $n_1 = n_2 = n_3 = n_4$. 

In what follows, we will provide, to one-loop order, the appropriate conversion factor to the $\MSbar$ scheme for all the above choices; it is given by:
\be
C_\Gamma^{{\rm GIRS},\MSbar} \equiv \frac{Z_\Gamma^{{\rm DR}, \MSbar}}{Z_\Gamma^{{\rm DR}, {\rm GIRS}}} = \frac{Z_\Gamma^{{\rm LR}, \MSbar}}{Z_\Gamma^{{\rm LR}, {\rm GIRS}}}.
\ee 
Any GF computed nonperturbatively and renormalized (also nonperturbatively) according to any of the above choices should, upon conversion to the $\MSbar$ scheme, lead to renormalized GFs which coincide, regardless of the prescription used; this then provides a very strong consistency check of the nonperturbative results. The ultimate selection of a prescription will depend on the statistical errors involved in each case.
  
For completeness, we note that there is another variant of GIRS given in~\cite{Tomii:2018zix}; in this approach, the average of the position vector $(x-y)$ is taken over the 3-dimensional surface of a hypersphere with radius $|x-y|$, centered at the origin.

\subsection{Tree-level order}
\label{treelevelorder}

\begin{figure}[h]
\centering
\epsfig{file=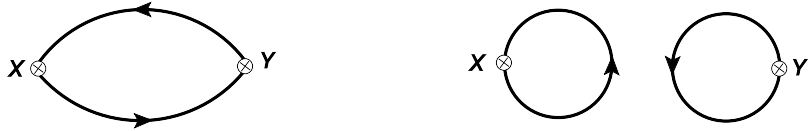,scale=0.65}
\caption{Tree-level Feynman diagrams contributing to the expectation value $\langle {\cal O}_{X}(x) {\cal O}_{Y}(y)\rangle$. A solid line represents quarks.
  }
\label{DRtree}
\end{figure}

The first step in our perturbative procedure is to calculate the tree-level value of the GF, $G(x-y) \equiv \langle\mathcal{O}_{X}(x)\mathcal{O}_{Y}(y) \rangle$, using dimensional regularization. This simple exercise serves to explain the procedure applied beyond tree level. The Feynman diagrams contributing to this expectation value are shown in Fig.~\ref{DRtree}. 
However, since we will consider the operators $\mathcal{O}_{X}$ and $\mathcal{O}_{Y}$ to be normal ordered, we discard the second diagram of Fig.~\ref{DRtree}. We are interested in a mass-independent scheme and thus all quark masses are set to zero. Then the tree-level contribution takes the following form:
\begin{equation}
    G^{{\rm tree}}(x-y)  =  N_c \int{\frac{d^dp}{(2\pi)^d}}\int{\frac{d^dk}{(2\pi)^d}}e^{i(k-p)(x-y)}\frac{1}{p^2\:k^2}\Tr(\slashed{p} Y\slashed{k} X),
\label{Gtree}
\end{equation}
where $N_c$ is the number of colors. Integrating over the momenta $p$, $k$, the resulting expression is:
\begin{equation}
    G^{{\rm tree}}(x-y) = \frac{N_c \ {(\Gamma(2-\varepsilon))}^2}{4 \pi^{4-2\varepsilon}} \frac{(x_{\mu}-y_{\mu}) \ (x_{\nu}-y_{\nu})}{{((x-y)^2)}^{4-2 \varepsilon}} \ \Tr(\gamma_\mu Y\gamma_\nu X).\\
 \label{Gtree1}
\end{equation}
where a summation over repeated indices $\mu$, $\nu$ is understood. One should observe that when $X$, $Y$ transform differently under rotations and parity, Eq. \eqref{Gtree1} gives zero. The results\footnote{The different definitions of $\gamma_5$ give identical results at tree level, since only the first four components of the vectors $x$ and $y$ can be nonzero.} for the nonzero cases are listed in Table~\ref{Tab:Tca}, in terms of an overall factor: $\mathcal{N} (x-y) \equiv N_c \ {(\Gamma(2-\varepsilon))}^2 / (4 \pi^{4-2\varepsilon}{((x-y)^2)}^{4- 2 \varepsilon} )$:
\begin{table}[H]
\centering
\begin{tabular}{M{2.5cm} || M{2.5cm} || M{9cm} N } 
\hline
\hline
& & & \\
$\mathbf{X}$ & $\mathbf{Y}$ &  $\bf{G^{{\rm tree}}(z) / \mathcal{N} (z)}$&\\[5pt] \hline
& & & \\
$1$ & $1$ & $4 z^2$ &\\[7pt] \hline
& & & \\
$\gamma_5$ & $\gamma_5$ & $-4 z^2$ &\\[7pt] \hline
& & & \\
$\gamma_{\nu_1}$ & $\gamma_{\nu_2}$ & $8 z_{\nu_1} z_{\nu_2} -4 \ \delta_{\nu_1 \nu_2} z^2$ &\\[7pt] \hline
& & & \\
$\gamma_5\gamma_{\nu_1} $ & $\gamma_5\gamma_{\nu_2} $ & $8 z_{\nu_1} z_{\nu_2} -4 \ \delta_{\nu_1 \nu_2} z^2$ &\\[7pt] \hline
& & & \\
$\sigma_{\nu_1 \nu_2}$ & $\sigma_{\nu_3 \nu_4}$ & $ 8 \ (\delta_{\nu_1 \nu_3} z_{\nu_2} z_{\nu_4} - \ \delta_{\nu_1 \nu_4} z_{\nu_2} z_{\nu_3} - \ \delta_{\nu_2 \nu_3} z_{\nu_1} z_{\nu_4} + \ \delta_{\nu_2 \nu_4} z_{\nu_1} z_{\nu_3}) -4 (\delta_{\nu_1 \nu_3} \delta_{\nu_2 \nu_4} - \delta_{\nu_1 \nu_4} \delta_{\nu_2 \nu_3}) \ z^2$ &\\[15pt] \hline
\hline
\end{tabular}
 \caption{Values of the tree-level contributions to the Green's functions
 $\big\langle\mathcal{O}_{X}(x)\mathcal{O}_{Y}(y)\big\rangle$ in $4-2\varepsilon$ dimensions in terms of an overall factor of $\mathcal{N} (z) \equiv N_c\: {(\Gamma(2-\varepsilon))}^2 / (4 \pi^{4-2\varepsilon}{(z^2)}^{4 - 2\varepsilon} ) $, where $z \equiv x-y$.} \label{Tab:Tca}
\end{table}

In order to apply t-GIRS, i.e. to integrate over spatial components, we consider all $8$ possibilities for $X$, $Y$ involving both time-like and space-like directions of Dirac matrices: $\openone,\;\gamma_5,\;\gamma_t,\;\gamma_i,\;\gamma_5\gamma_t,\;\gamma_5\gamma_i,\:\gamma_t\gamma_i,\;\gamma_i\gamma_j$, where $i \ne j \ne t \ne i$. The only nonvanishing contribution under integration over the spatial components, stems from the case $X = Y$ and $\mu=\nu$ in Eq.~\eqref{Gtree1}. The results for $G^{{\rm tree}} (\vec{x},t)$ in 4 dimensions, after integrating over the spatial components $\vec{x}$, are given, in terms of an overall factor of $N_c / (\pi^2 {|t|}^3)$, in Table \ref{Tab:Tca2}. We note that the integrated GFs for $\gamma_t$ and $\gamma_5 \gamma_t$ vanish; the consequence of this fact for the corresponding operators will be discussed in the following subsection.

\begin{table}[H]
\centering
\begin{tabular}{M{2.5cm} || M{5cm} N } 
\hline
\hline
& \\
$\mathbf{X} = \mathbf{Y}$ &  $\mathbf{ \int d^3} \boldsymbol{\vec{x}} \ \mathbf{G^{{\rm \textbf{tree}}}(}\boldsymbol{\vec{x}}\mathbf{, t)} / [N_c / (\pi^2 {|t|}^3)]$&\\[5pt] \hline
& \\
$1$ & $\phantom{+}1 / 4$ &\\[7pt] \hline
& \\
$\gamma_5$ & $-1 / 4$ &\\[7pt] \hline
& \\
$\gamma_{t}$ & $\phantom{+}0$ &\\[7pt] \hline
& \\
$\gamma_{i}$ & $- 1 / 6$ &\\[7pt] \hline
& \\
$\gamma_5\gamma_{t} $ & $\phantom{+}0$ &\\[7pt] \hline
& \\
$\gamma_5\gamma_{i} $ & $- 1 / 6$ &\\[7pt] \hline
& \\
$\sigma_{t i}$ & $\phantom{+}1 / 12$ &\\[7pt] \hline
& \\
$\sigma_{i j}$ & $- 1 / 12$ &\\[7pt] \hline
\hline
\end{tabular}
 \caption{Values of the tree-level contributions to the Green's functions
 $\big\langle\mathcal{O}_{X}(\vec{x},0)\mathcal{O}_{Y}(\vec{0},t)\big\rangle$ in 4 dimensions, in terms of an overall factor of $N_c / (\pi^2 {|t|}^3)$, after integrating over $\vec{x}$.} \label{Tab:Tca2}
\end{table}

\subsection{One-loop order}
\label{onelooporder}

\begin{figure}[h]
\centering
\epsfig{file=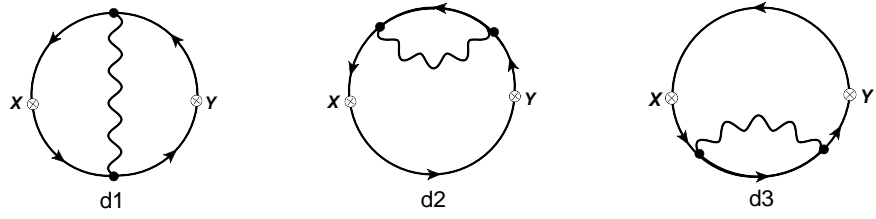,scale=0.7}
\caption{One-loop order Feynman diagrams contributing to the expectation value $\langle {\cal O}_{X}(x) {\cal O}_{Y}(y)\rangle$. A wavy (solid) line represents gluons (quarks).
  }
\label{DR1loop}
\end{figure}

At one-loop level, the Feynman diagrams which contribute to the GFs $G (x-y)$ are given in Fig.~\ref{DR1loop}.

\hspace{-0.5cm} The corresponding contributions are shown below: \\
\\
Diagram 1:\\
\begin{equation}
d_1 = \int\frac{d^dk}{(2\pi)^d}   \int\frac{d^dp}{(2\pi)^d}\int\frac{d^dq}{(2\pi)^d}\; \frac{g_B^2}{2}(N_c^2-1)e^{ik(x-y)}\bigg[\frac{1}{(p-q)^2p^2(p-k)^2q^2(q-k)^2}Tr(X\slashed{p}\gamma_\mu\slashed{q}Y(\slashed{q}-\slashed{k})\gamma_\mu(\slashed{p}-\slashed{k}))\bigg],
\label{diagram1}
\end{equation}\\
Diagram 2:\\
\begin{equation}
d_2 = \int\frac{d^dk}{(2\pi)^d}   \int\frac{d^dp}{(2\pi)^d}\int\frac{d^dq}{(2\pi)^d}\; \frac{g_B^2}{2}(N_c^2-1)e^{ik(x-y)}\bigg[\frac{1}{(p-q)^2(p^2)^2(p-k)^2q^2}Tr(X\slashed{p}\gamma_\mu\slashed{q}\gamma_\mu\slashed{p}Y(\slashed{p}-\slashed{k}))\bigg],
\label{diagram2}
\end{equation}\\
Diagram 3:\\ 
\begin{equation}
d_3 = \int\frac{d^dk}{(2\pi)^d}   \int\frac{d^dp}{(2\pi)^d}\int\frac{d^dq}{(2\pi)^d}\; \frac{g_B^2}{2}(N_c^2-1)e^{ik(x-y)}\bigg[\frac{1}{(p-q)^2(p^2)^2(p-k)^2q^2}Tr(X(\slashed{p}-\slashed{k})Y\slashed{p}\gamma_\mu\slashed{q}\gamma_\mu\slashed{p})\bigg].
\label{diagram3}
\end{equation}\\ 
Note that for one-loop calculations, the bare and renormalized coupling constants differ only by the factor $\mu^{-\varepsilon}$ (see Eq. \eqref{RenDef}), as only the tree-level value of $Z_g = 1 + \mathcal{O} (g^2)$ contributes in this case (i.e., $Z_g \rightarrow 1$).

\begin{figure}[h]
\centering
\epsfig{file=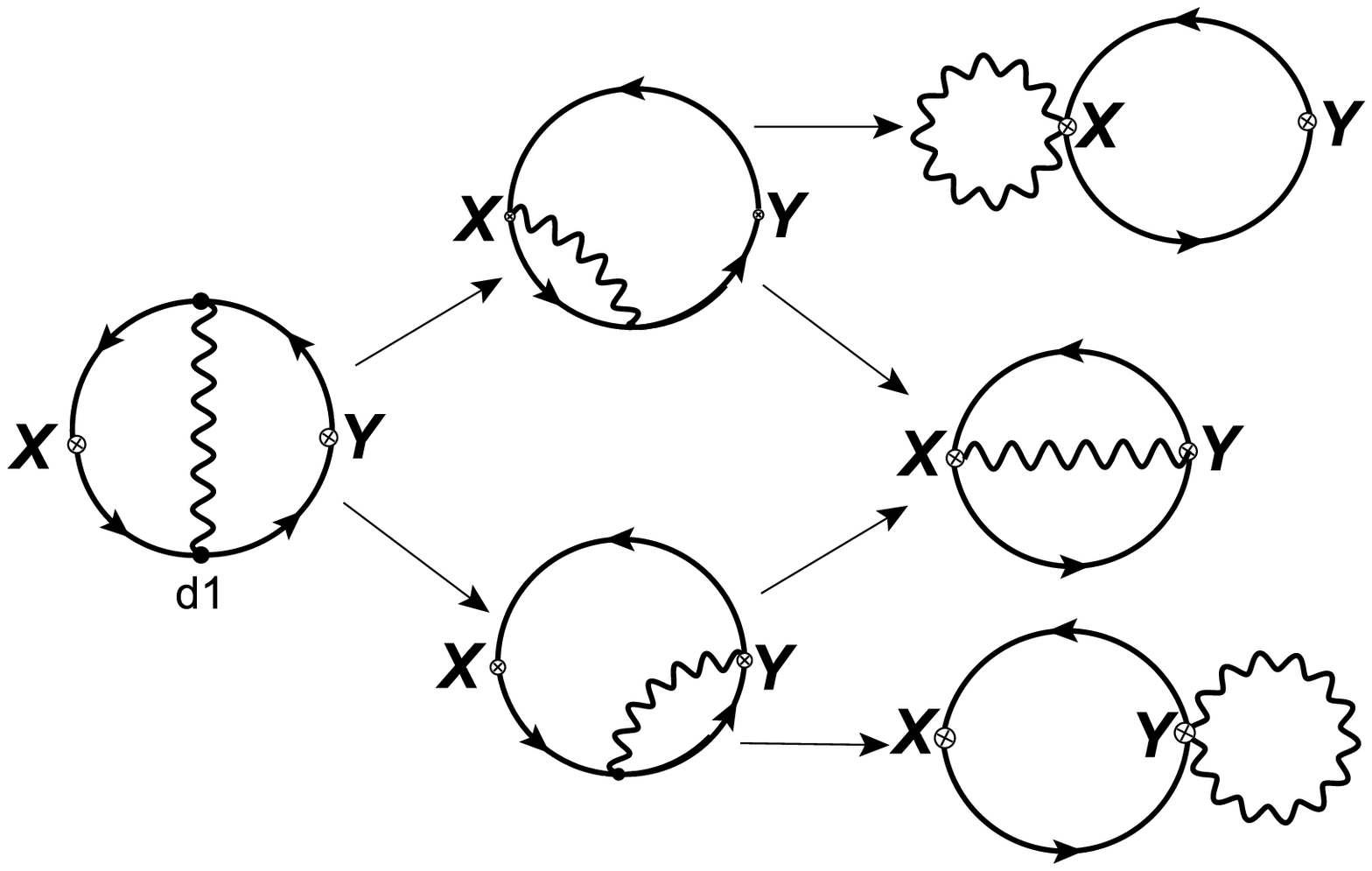,scale=0.35}\;
\epsfig{file=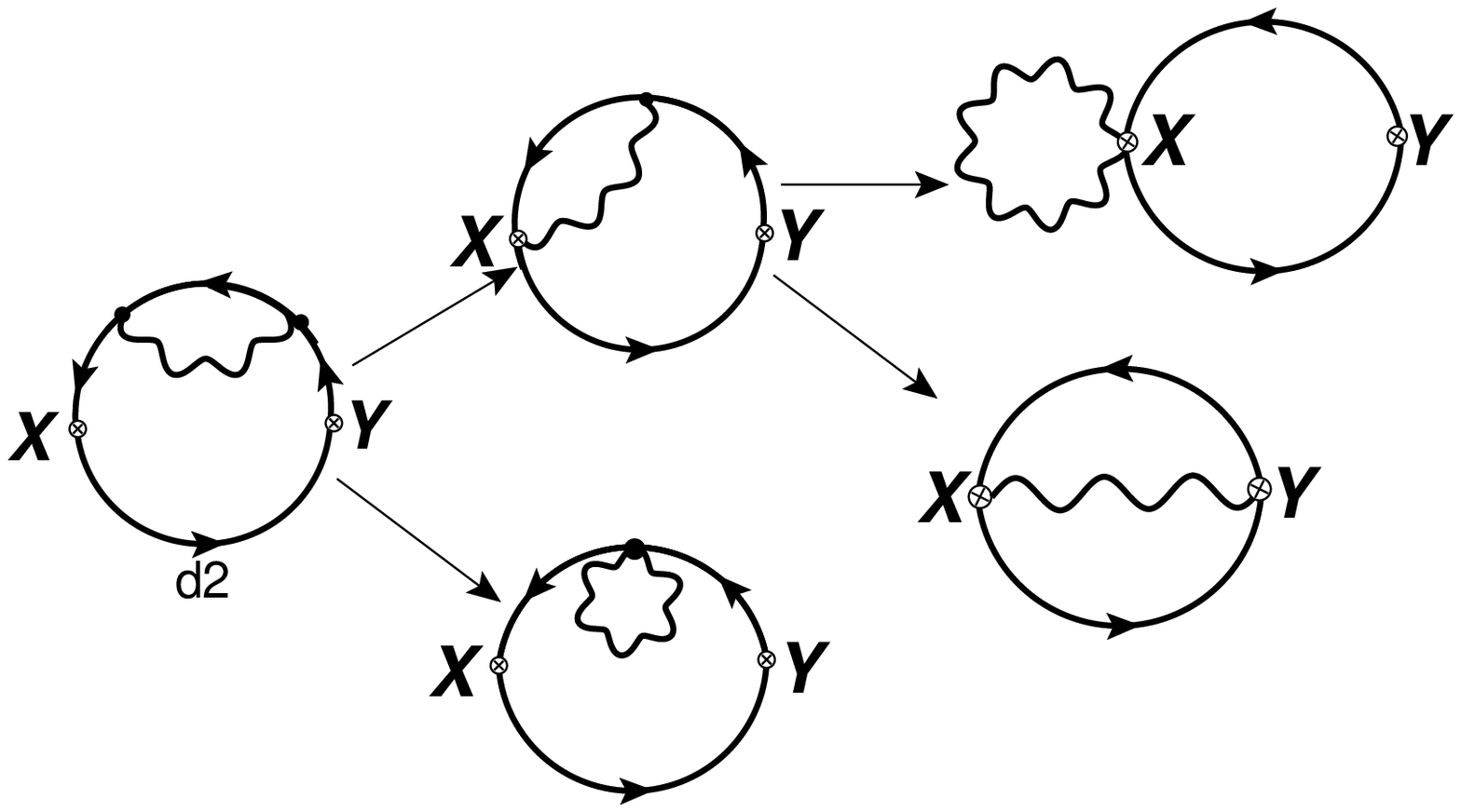,scale=0.35}\;
\epsfig{file=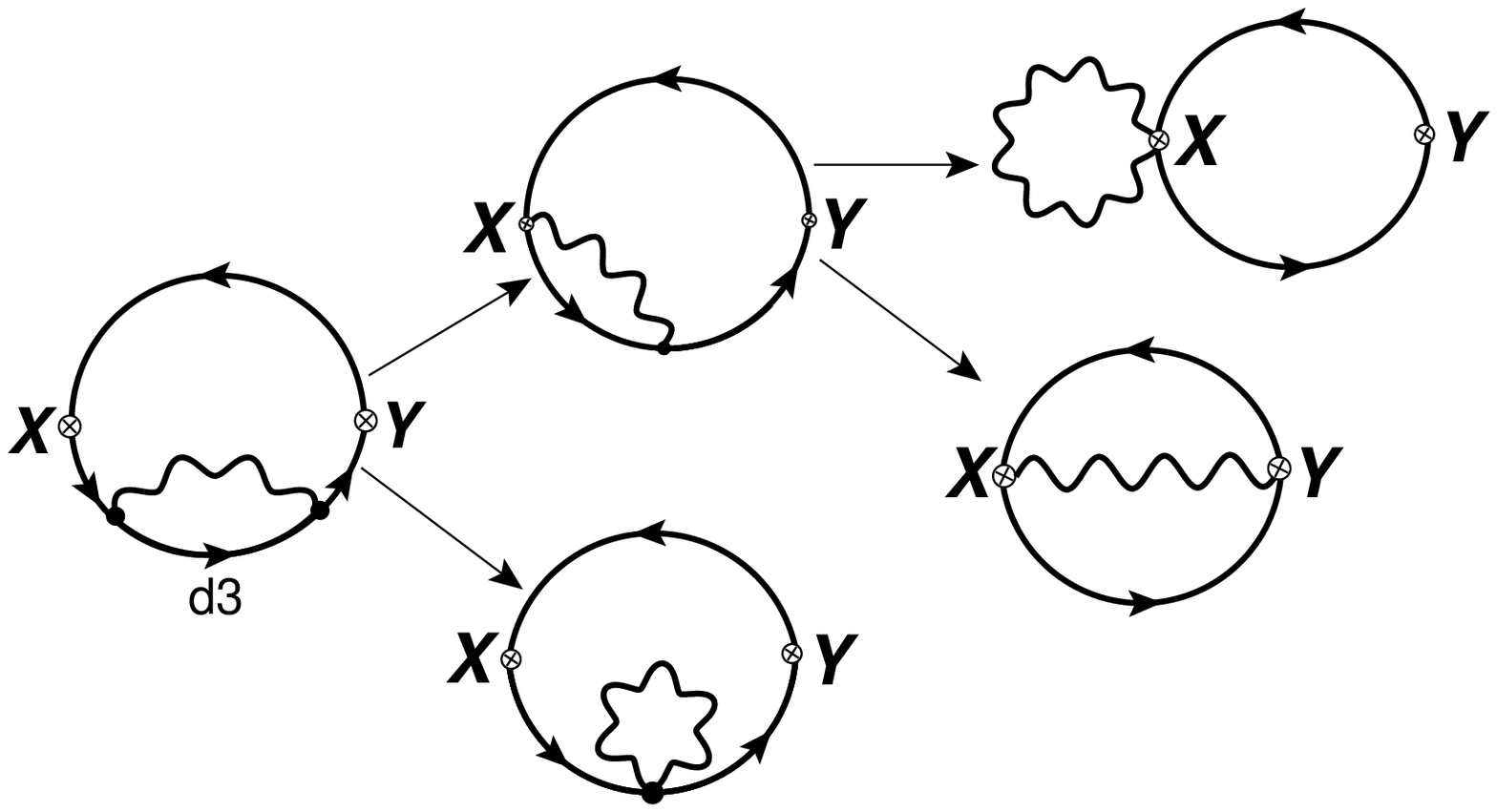,scale=0.35}
\caption{Diagrammatic representation of the application of the Ward Identity on the three one-loop diagrams.}
\label{GI_DR}
\end{figure}

Our next step is to verify that the one-loop contribution is indeed gauge-invariant, i.e., to verify that terms depending on the gauge parameter\footnote{$\beta = 0 \ (1)$ corresponds to the Feynman (Landau) gauge.}, $\beta$, cancel out when we sum the three diagrams. To this end, we may use the following Ward-like identity:
\be
\frac{\slashed{k}}{\slashed{p} \ (\slashed{p}+\slashed{k})} = \frac{1}{\slashed{p}}-\frac{1}{\slashed{p}+\slashed{k}}
\label{GInva}
\ee
If we implement this identity on the vertex of a diagram, the initial diagram gets split into two new ones that have a fermion propagator removed. The double application of the above identity on the three one-loop diagrams is given diagrammatically in Fig.~\ref{GI_DR}.
The tadpole diagrams lead to a scaleless expression, which therefore vanishes in dimensional regularization. The remaining contributions ($\beta$-terms) cancel against each other when summed. We can thus focus on the Feynman gauge.

Recall that the massless tree-level GFs were proportional to a trace of the form $ {\rm Tr}( \slashed{z} Y \slashed{z} X)$. The divergent part of the one-loop contribution is expected to contain the same traces. After some Dirac trace algebra, it turns out that the finite parts of GFs are proportional to the tree level as well. In order to determine the conversion factors between GIRS and $\MSbar$, we must compare one-loop GFs to the corresponding tree-level ones. However, when integrating over spatial components, the GFs containing the operators with $\gamma_t$ and $\gamma_5\gamma_t$ vanish at tree-level and one-loop order; thus, this type of GF cannot be used to evaluate the renormalization factors of these particular operators. Nevertheless, we can adopt the natural definition that these renormalization factors be the same as those of $\gamma_i$ and $\gamma_5\gamma_i$, respectively.

Higher-loop contributions involve continuum integrands over more than three momenta, including exponentials in the numerator; these make the calculation process more complicated. Note, however, that the presence of exponentials in two-point functions amounts to a simple Fourier transform, once integrals over inner momenta have been performed; such integrals involve only one external momentum and massless propagators, and they
have been studied in various contexts in the literature to higher loop order. In this study, we limit ourselves up to one-loop computations. For a four-loop evaluation, see Ref.~\cite{Chetyrkin:2010dx}.

The procedure of formulating a gauge-invariant renormalization scheme entails performing perturbative calculations in the continuum, while the necessary lattice calculations can be performed in a completely nonperturbative way. Still, calculations in lattice perturbation theory can be used to check the validity of nonperturbative methods. Also, a perturbative calculation may be employed in order to reduce cutoff effects present in the nonperturbative estimates. While perturbative calculations are easier to implement in the continuum (and also unavoidable by the very nature of the $\MSbar$ scheme), they become exceedingly complicated on the lattice, and consequently, calculations beyond two loops are practically unfeasible\footnote{The only three-loop calculations on the lattice existing thus far regard ``vacuum'' diagrams, that is diagrams without external lines and momenta~\cite{Alles:1998is,Panagopoulos:2006ky,Athenodorou:2007hi}. In addition, stochastic perturbation theory has been carried out to higher loops~\cite{Brambilla:2013sua}.}. In addition, even some two-loop calculations become prohibitive for some ``improved'' actions, such as the ones used in many large-scale simulations nowadays. Thus, in practice, lattice results are typically limited to one loop, and this can lead to large systematic errors. In GIRS, even the one-loop calculation is not trivial since, as mentioned in Sec. \ref{Introduction}, ``$n$-loop'' (i.e., order $g^{2n}$) calculations involve $(n+1)$-loop Feynman diagrams. Also, the presence of exponentials in Feynman integrals makes their computation more complex. A method for calculating similar integrals on the lattice can be found in Ref.~\cite{Constantinou:2017sej}, where the renormalization of nonlocal operators is studied; however, in that work only one-loop Feynman diagrams were considered.  

\subsection{Results}
\label{RI}

In this subsection, we present our results (up to one loop) on the bare GFs $G (x-y)$, as well as the conversion factors between all the variants of GIRS and $\MSbar$ scheme. As mentioned above, we employ both NDR and HV prescriptions; HV is more useful for comparison with experimental determinations and phenomenological estimates, while NDR is applied for comparison with previous calculations. The one-loop conversion factor between NDR and HV prescriptions can be extracted from our results. 

Our resulting expressions for the five nonvanishing bare GFs are ($z \equiv x-y$):
\bea
\langle {\cal O}^{\rm DR}_{\openone}(x) \ {\cal O}^{\rm DR}_{\openone}(y) \rangle &=& \frac{N_c \ {(\Gamma (2-\varepsilon))}^2}{\pi^{4 - 2 \varepsilon} \ (z^2)^{3 - 2 \varepsilon}} \Bigg[ 1 + \frac{g_\MSbar^2 \ C_F}{16 \pi^2} \left(\frac{6}{\varepsilon} + 2 + 6 \ln (\bar \mu ^2 z^2) + 12 \gamma_E - 12 \ln(2) \right) \nonumber \\
&& \qquad \qquad \qquad \qquad + \ \mathcal{O} (\varepsilon^1, g_\MSbar^2) \ + \ \mathcal{O} (g_\MSbar^4) \Bigg], \label{GS}\\
\langle {\cal O}^{\rm DR}_{\gamma_5}(x) \ {\cal O}^{\rm DR}_{\gamma_5}(y) \rangle &=& - \langle {\cal O}^{\rm DR}_{\openone}(x) {\cal O}^{\rm DR}_{\openone}(y) \rangle - 16 \, {\rm hv} \, \frac{N_c \ {(\Gamma (2-\varepsilon))}^2}{\pi^{4 - 2 \varepsilon} \ (z^2)^{3 - 2 \varepsilon}} \frac{g_\MSbar^2 \ C_F}{ 16 \pi^2} \ + \ \mathcal{O} (g_\MSbar^4), \\
\langle {\cal O}^{\rm DR}_{\gamma_{\nu_1}}(x) \ {\cal O}^{\rm DR}_{\gamma_{\nu_2}}(y) \rangle &=& \frac{N_c \ {(\Gamma (2-\varepsilon))}^2}{\pi^{4 - 2 \varepsilon} \ (z^2)^{4 - 2 \varepsilon}} \Big(2 \ z_{\nu_1} \ z_{\nu_2} - \delta_{\nu_1 \nu_2} \ z^2  \Big)  \left( 1 + 3 \frac{g_\MSbar^2 \ C_F}{16 \pi^2} \ + \ \mathcal{O} (\varepsilon^1, g_\MSbar^2) \ + \ \mathcal{O} (g_\MSbar^4) \right), \ \\
\langle {\cal O}^{\rm DR}_{\gamma_5 \gamma_{\nu_1}}(x) \ {\cal O}^{\rm DR}_{\gamma_5 \gamma_{\nu_2}}(y) \rangle &=& \langle {\cal O}^{\rm DR}_{\gamma_{\nu_1}}(x) {\cal O}^{\rm DR}_{\gamma_{\nu_2}}(y) \rangle + 8 \, {\rm hv} \, \frac{N_c \ {(\Gamma (2-\varepsilon))}^2}{\pi^{4 - 2 \varepsilon} \ (z^2)^{4 - 2 \varepsilon}} \Big(2 \ z_{\nu_1} \ z_{\nu_2} - \delta_{\nu_1 \nu_2} \ z^2 \Big) \frac{g_\MSbar^2 \ C_F}{16 \pi^2} \ + \ \mathcal{O} (g_\MSbar^4), \ \\
\langle {\cal O}^{\rm DR}_{\sigma_{\nu_1 \nu_2}}(x) \ {\cal O}^{\rm DR}_{\sigma_{\nu_3 \nu_4}}(y) \rangle &=&  \frac{N_c \ {(\Gamma (2-\varepsilon))}^2}{\pi^{4 - 2 \varepsilon} \ (z^2)^{4 - 2 \varepsilon} } \Big[2 \ (\delta_{\nu_1 \nu_3} \ z_{\nu_2} \ z_{\nu_4} - \delta_{\nu_1 \nu_4} \ z_{\nu_2} \ z_{\nu_3} - \delta_{\nu_2 \nu_3} \ z_{\nu_1} \ z_{\nu_4} + \delta_{\nu_2 \nu_4} \ z_{\nu_1} \ z_{\nu_3}) \nonumber\\
&& \qquad \qquad \qquad \quad - (\delta_{\nu_1 \nu_3} \ \delta_{\nu_2 \nu_4} - \delta_{\nu_1 \nu_4} \ \delta_{\nu_2 \nu_3}) z^2 \Big] \times \nonumber\\
&& \qquad \qquad \qquad \quad \Bigg[ 1 + \frac{g_\MSbar^2 \ C_F}{ 16 \pi^2} \left(- \frac{2}{\varepsilon} + 6 - 2 \ln (\bar \mu^2 z^2)-4\gamma_E  + 4 \ln (2) \right) \nonumber \\
&& \qquad \qquad \qquad \qquad + \ \mathcal{O} (\varepsilon^1, g_\MSbar^2) \ + \ \mathcal{O} (g_\MSbar^4) \Bigg], \qquad \label{GT}
\eea
where $C_F=(N_c^2-1)/(2\,N_c)$ is the quadratic Casimir operator in the fundamental representation and ${\rm hv} = 0 \ (1)$ for the NDR (HV) prescription of $\gamma_5$. Our results agree with Ref.~\cite{Gimenez:2004me} (in the limit $\varepsilon \rightarrow 0$)\footnote{Up to possible typos: An overall minus sign is missing in the pseudoscalar case. Furthermore, a sign must be altered in the definition of the parameter $\hat{\varepsilon}$, as follows: $1/\hat{\varepsilon} \equiv 1/\varepsilon + \ln (4 \pi) - \gamma_E$.}, in the case ${\rm hv} = 0$ and $N_c = 3$. These results can be used to derive the renormalization factors in $\MSbar$ and in any variant of GIRS. In particular, the $\MSbar$-renormalized GFs are the same as the above, once the $1 / \varepsilon$ poles are removed and the na{\"i}ve limit $\varepsilon \rightarrow 0$ is taken in the remaining terms. The vector and axial vector cases are free of poles, as expected.

For another crosscheck of our results, we extract the multiplicative renormalization factors $Z_{\Gamma}^{{\rm DR}, \MSbar}$, using the following relation\footnote{As usual, perturbative corrections in $Z_\Gamma^{{\rm DR},\MSbar}$ may only be proportional to inverse powers of $\varepsilon$, with coefficients chosen in a way as to give a well-defined limit to the right-hand side of Eq.~\eqref{MSbar}.}: 
\begin{equation}
     \big\langle\mathcal{O}^\MSbar_X(x)\mathcal{O}^\MSbar_Y(y)\big\rangle =  Z_X^{{\rm DR}, \MSbar} Z_Y^{{\rm DR}, \MSbar} \ \big\langle\mathcal{O}^{\rm DR}_X(x)\mathcal{O}^{\rm DR}_Y(y)\big\rangle|_{\varepsilon \to \ 0}.
     \label{MSbar}
\end{equation}
The renormalization factors can be read directly from the bare GFs (Eqs. (\ref{GS} - \ref{GT})):
\bea
Z_S^{DR,\MSbar} &=& 1 \ - \ \frac{g_\MSbar^2 \ C_F}{16 \pi^2} \frac{3}{\varepsilon} \ + \ {\cal O} (g_\MSbar^4),\\
Z_P^{DR,\MSbar} &=& 1 \ - \ \frac{g_\MSbar^2 \ C_F}{16 \pi^2} \frac{3}{\varepsilon} \ + \ {\cal O} (g_\MSbar^4),\\
Z_V^{DR,\MSbar} &=& 1 \ + \ {\cal O} (g_\MSbar^4),\\
Z_A^{DR,\MSbar} &=& 1 \ + \ {\cal O} (g_\MSbar^4),\\
Z_T^{DR,\MSbar} &=& 1 \ + \ \frac{g_\MSbar^2 \ C_F}{16 \pi^2} \frac{1}{\varepsilon} \ + \ {\cal O} (g_\MSbar^4),
\eea
where we use the notation \{S,P,V,A,T\} for \{scalar, pseudoscalar, vector, axial-vector, tensor\} operators. These factors are in agreement with well-known results in the literature (\cite{Gracey:2003yr} and references therein).

Applying the condition of Eq. \eqref{GIRScond} to the resulting expressions for the bare GFs [Eqs. (\ref{GS} - \ref{GT})], we extract the conversion factors between GIRS and $\MSbar$ schemes:
\bea
C_{S}^{{\rm GIRS},\MSbar} &=& 1 \ + \ \frac{g_\MSbar^2 \ C_F}{16 \pi^2} \left[1 + 3 \ln (\bar \mu ^2 \bar{z}^2) + 6 \gamma_E - 6 \ln(2) \right] \ + \ \mathcal{O} (g_\MSbar^4), \\
C_{P}^{{\rm GIRS},\MSbar} &=& 1 \ + \ \frac{g_\MSbar^2 \ C_F}{16 \pi^2} \left[1 + 3 \ln (\bar \mu ^2 \bar{z}^2)+ 6 \gamma_E - 6 \ln(2) + 8 \, {\rm hv} \right] \ + \ \mathcal{O} (g_\MSbar^4), \\
C_{V}^{{\rm GIRS},\MSbar} &=& 1 \ + \ \frac{g_\MSbar^2 \ C_F}{16 \pi^2} \left(\frac{3}{2} \right) \ + \ \mathcal{O} (g_\MSbar^4), \\
C_{A}^{{\rm GIRS},\MSbar} &=& 1 \ + \ \frac{g_\MSbar^2 \ C_F}{16 \pi^2} \left( \frac{3}{2} + 4\,{\rm hv} \right) \ + \ \mathcal{O} (g_\MSbar^4), \\
C_{T}^{{\rm GIRS},\MSbar} &=& 1 \ + \ \frac{g_\MSbar^2 \ C_F}{16 \pi^2} \left[3 - \ln (\bar \mu^2 \bar{z}^2)-2\gamma_E  + 2 \ln (2) \right] \ + \ \mathcal{O} (g_\MSbar^4).
\eea
Note that the one-loop results of the bare GFs are proportional to the tree-level ones, and therefore only the length (not the orientation) of $x-y$ is relevant as a renormalization scale. Integrating Eqs. (\ref{GS} - \ref{GT})) over spatial components and applying the condition of Eq. \eqref{RC2}, we also extract the conversion factors between t-GIRS and $\MSbar$ schemes. As previously mentioned, the integration over spatial components separates further these cases into $8$ possibilities ($S, P, V_t, V_i, A_t, A_i, T_{ti}, T_{ij}$) which depend on whether $(x_\mu-y_\mu)$ is temporal or not, and they correspond to $X=Y=\{1,\;\gamma_5,\;\gamma_t,\;\gamma_i,\;\gamma_5\gamma_t,\;\gamma_5\gamma_i,\:\gamma_t\gamma_i,\;\gamma_i\gamma_j\}$. Two of them ($V_t, A_t$) give a vanishing contribution, both at tree level, and one loop; however, it is natural to impose that $V_t (A_t)$ has the same renormalization factor as $V_i (A_i)$. Thus, below we present results for the remaining $6$ operators. Note that for extracting the correct one-loop renormalization factors in GIRS, it is essential to include ${\cal O} (\varepsilon^1)$ terms of the tree-level GFs $G^{\rm tree} (x-y)$; we recall that such terms are also necessary in the evaluation of the $\MSbar$-renormalized GFs. The conversion factors are:  
\bea
C_{S}^{{\rm t-GIRS},\MSbar} &=& 1 \ + \ \frac{g_\MSbar^2 \ C_F}{16 \pi^2} \left(-\frac{1}{2} + 6 \ln (\bar{\mu} \bar{t}) +6 \gamma_E\right) + \mathcal{O} (g_\MSbar^4),\\
C_{P}^{{\rm t-GIRS},\MSbar} &=& 1 \ + \ \frac{g_\MSbar^2 \ C_F}{16 \pi^2}\left(-\frac{1}{2} + 6 \ln (\bar{\mu} \bar{t}) + 6 \gamma_E + 8 \, {\rm hv} \right) + \mathcal{O} (g_\MSbar^4),\\
C_{V_i}^{{\rm t-GIRS},\MSbar} &=& 1 \ + \ \frac{g_\MSbar^2 \ C_F}{16 \pi^2} \left(\frac{3}{2}\right) + \mathcal{O} (g_\MSbar^4),\\
C_{A_i}^{{\rm t-GIRS},\MSbar} &=& 1 \ + \ \frac{g_\MSbar^2 \ C_F}{16 \pi^2} \left(\frac{3}{2} + 4\,{\rm hv}\right) + \mathcal{O} (g_\MSbar^4),\\
C_{T_{ti}}^{{\rm t-GIRS},\MSbar} &=& 1 \ + \ \frac{g_\MSbar^2 C_F}{16 \pi^2}\left(\frac{25}{6} -2 \ln(\bar{t} \bar \mu) -2 \gamma_E \right) + \mathcal{O} (g_\MSbar^4),\\
C_{T_{ij}}^{{\rm t-GIRS},\MSbar} &=& 1 \ + \ \frac{g_\MSbar^2 \ C_F}{16 \pi^2}\left(\frac{25}{6} -2 \ln(\bar{t} \bar \mu) -2 \gamma_E \right) + \mathcal{O} (g_\MSbar^4).
\eea

\section{Renormalization and mixing of the quark and gluon energy-momentum tensor operators in GIRS}
\label{Renormalization and mixing of the quark and gluon energy-momentum tensor operators in GIRS}

\subsection{Details of the Calculation}

The gauge-invariant part of the QCD traceless symmetric energy-momentum tensor (EMT)~\cite{Freedman:1974gs} contains two flavor-singlet operators, a gluonic $\mathcal{O}_1^{\mu \nu}$ and a fermionic $\mathcal{O}_2^{\mu \nu}$:
\begin{eqnarray}
\mathcal{O}_1^{\mu \nu} &=& 2 {\rm Tr} [F^{\mu \rho} F^{\nu \rho}] - \frac{1}{d} \delta^{\mu \nu} 2 {\rm Tr} [F^{\rho \sigma} F^{\rho \sigma}], \label{O1} \\
\mathcal{O}_2^{\mu \nu} &=& \sum_{f=1}^{N_f} \left[ \frac{1}{2} \left( \bar{\psi}_f \gamma^\mu \overleftrightarrow{D}^\nu \psi_f + \bar{\psi}_f \gamma^\nu \overleftrightarrow{D}^\mu \psi_f \right) - \frac{1}{d} \delta^{\mu \nu} \left( \bar{\psi}_f \gamma^\rho \overleftrightarrow{D}^\rho \psi_f \right) \right], \label{O2}
\end{eqnarray}
where $F^{\mu \nu} \equiv \partial^\mu A^\nu - \partial^\nu A^\mu + i g [A^\mu, A^\nu]$, $\overleftrightarrow{D}^\mu \equiv ( \overrightarrow{D}^\mu - \overleftarrow{D}^\mu ) / 2$, $\overrightarrow{D}^\mu \equiv \overrightarrow{\partial}^\mu + i g A^\mu$, and $\overleftarrow{D}^\mu \equiv \overleftarrow{\partial}^\mu - i g A^\mu$; a summation over repeated indices is implied. These two operators are involved in the structure functions in nucleons~\cite{Ji:1995sv, Ji:1996ek, Radyushkin:1997ki, Ji:1998pc}: the gluon operator appears in the leading-twist approximation of the gluon parton distribution function, while the fermion operator is related to the unpolarized quark parton distribution function. Furthermore, their matrix elements are directly related to the gluon and quark average momentum fraction of a nucleon state~\cite{Ji:1998pc, Horsley:2012pz}. Also, these operators are connected to the anomalous magnetic moment of the muon~\cite{Aoyama:2020ynm}. 

A proper renormalization of the above gluon and fermion operators is required, before one can relate their matrix elements, as extracted from numerical simulations, to physical observables. A difficulty in calculating these renormalization factors is that mixing is present; these operators mix among themselves as they have the same transformations under Euclidean rotational (or hypercubic, on the lattice) symmetry. They also mix with other operators, including gauge-variant operators (BRST variations and operators which vanish by the equations of motion; see~\cite{Caracciolo:1991cp}). The latter include ghost and gauge-fixing terms, which are well-defined in perturbation theory, while their nonperturbative extensions are not obvious; thus, a nonperturbative study of such terms by compact lattice simulations is problematic. However, implementing a gauge-invariant renormalization scheme, such as GIRS, which involves only gauge-invariant GFs, the gauge-variant operators do not contribute in the renormalization process and thus, they are excluded. In our study, we consider only the nondiagonal elements ($\mu \neq \nu)$ of the above operators, which give a reduced set of operators under mixing on the lattice, containing only $\mathcal{O}_1^{\mu \nu}$ and $\mathcal{O}_2^{\mu \nu}$. Thus, the renormalization of the nondiagonal components of EMT operators entails the construction of a $2 \times 2$ mixing matrix, which relates the bare to the renormalized operators: 
\begin{equation}
\begin{pmatrix}
{\mathcal{O}_1^{\mu \nu}}^R \\
\\
{\mathcal{O}_2^{\mu \nu}}^R
\end{pmatrix} = 
\begin{pmatrix}
Z^{B, R}_{11} & Z^{B, R}_{12} \\
\\
Z^{B, R}_{21} & Z^{B, R}_{22}
\end{pmatrix} 
\begin{pmatrix}
{\mathcal{O}_1^{\mu \nu}}^B \\
\\
{\mathcal{O}_2^{\mu \nu}}^B
\end{pmatrix}.
\end{equation}  
   
\begin{figure}[h]
\centering
\epsfig{file=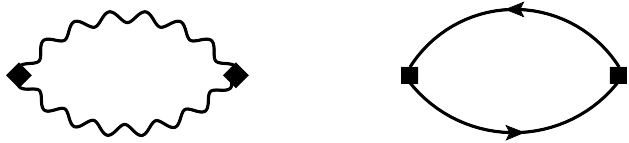,scale=0.5}
\caption{Diagrams which contribute to the tree-level Green's functions $G_{11}^{\nu_1 \nu_2; \nu_3 \nu_4}$ and $G_{22}^{\nu_1 \nu_2; \nu_3 \nu_4}$. A diamond (square) denotes insertion of $\mathcal{O}_1^{\mu \nu}$ ($\mathcal{O}_2^{\mu \nu}$).} \label{treeEMT}
\end{figure}

\begin{figure}[h]
\centering
\epsfig{file=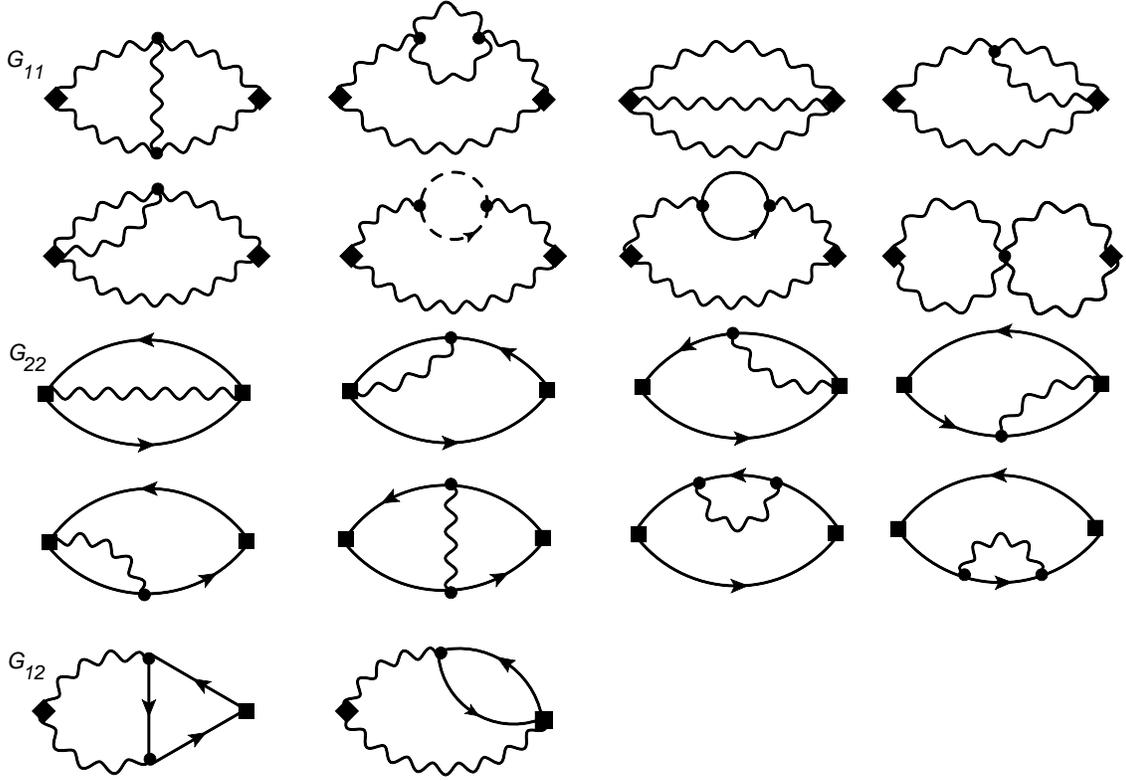,scale=0.9}
\caption{Diagrams which contribute to the one-loop Green's functions $G_{11}^{\nu_1 \nu_2; \nu_3 \nu_4}$, $G_{22}^{\nu_1 \nu_2; \nu_3 \nu_4}$ and $G_{12}^{\nu_1 \nu_2; \nu_3 \nu_4}=G_{21}^{\nu_1 \nu_2; \nu_3 \nu_4}$. Wavy (solid, dashed) lines represent gluons (quarks, ghosts). A diamond (square) denotes insertion of $\mathcal{O}_1^{\mu \nu}$ ($\mathcal{O}_2^{\mu \nu}$).} \label{oneloopEMT11}
\end{figure}

The calculation of the $2 \times 2$ mixing matrix requires a total of four conditions involving GFs of $\mathcal{O}_1^{\mu \nu}$ and $\mathcal{O}_2^{\mu \nu}$. Three different two-point GFs can be constructed by taking vacuum expectation values between the two mixing operators:
\begin{equation}
    G^{\nu_1 \nu_2; \nu_3 \nu_4}_{ij}(x-y)  \equiv  \big\langle\mathcal{O}^{\nu_1 \nu_2}_i(x) \ \mathcal{O}^{\nu_3 \nu_4}_j(y)\big\rangle, \quad (i,j) = (1,1), (1,2), (2,2),
\label{Gtree}
\end{equation}
where $\nu_1 \neq \nu_2$ and $\nu_3 \neq \nu_4$. By rotational (or just hypercubic) invariance: $G_{ij}^{\nu_1 \nu_2; \nu_3 \nu_4}(x-y) = G_{ji}^{\nu_1 \nu_2; \nu_3 \nu_4}(x-y) = G_{ji}^{\nu_3 \nu_4; \nu_1 \nu_2}(x-y)$. As it turns out, the two-point GFs $G_{ij}$ at one-loop level are proportional to the tree-level values of the diagonal elements $G_{ii}$, with a proportionality factor which is independent of the values of the Lorentz indices $\nu_i$; as a consequence, Eq. \eqref{Gtree} can lead to only three independent renormalization conditions. We calculate the above GFs up to one loop in dimensional regularization. The tree-level contributions come from Feynman diagrams shown in Fig.~\ref{treeEMT}, while the one-loop contributions stem from Feynman diagrams of Fig.~\ref{oneloopEMT11}. Note that $G_{12}^{\rm tree}=0$. 

From the above GFs we can get three renormalization conditions by requiring absence of poles (for $\MSbar$) or equality to the corresponding tree-level values in a renormalization position scale (for GIRS). These are not enough for fully determining the mixing matrix $Z_{ij}$ in a univocal way, in either $\MSbar$ or GIRS. In order to impose a fourth condition, we need to compute additional GFs; a most natural choice involves products of operators $\mathcal{O}^{\mu \nu}_i$ with lower-dimensional operators. The procedure, which is also alluded to in~\cite{Gracey:2009da}, has the form of a bootstrap: One starts by  renormalizing lowest dimensional operators (where no mixing issues are present), and then proceeds to renormalize operators of increasingly higher dimensionality by requiring finiteness (or some other normalization condition) in the GFs involving products of operators up to that dimensionality. In this case, the only available lower-dimensional gauge-invariant local operators are the fermion bilinears, studied in the previous section. 

The simplest GF that one might consider is a two-point function constructed from the product of an EMT operator $\mathcal{O}^{\mu \nu}_i$ and one fermion bilinear  ${\cal O}_{\Gamma}$ at two distinct spacetime points:
\begin{equation}
\langle \mathcal{O}_i^{\mu \nu} (x) \ \mathcal{O}_\Gamma (y) \rangle.
\end{equation}
However, such a GF vanishes for any choice of $\Gamma$ to all perturbative orders: the corresponding two-point GFs with a scalar, pseudoscalar or tensor operator give traces of an odd number of Dirac matrices (for massless fermions), while the GF with an axial vector gives traces containing one $\gamma_5$ and Dirac matrices with symmetrized indices; even the GF  with a vector operator $\langle \mathcal{O}^{\nu_1 \nu_2}_i (x) {\cal O}_{\gamma_{\nu_3}} (y) \rangle$ vanishes, since $\mathcal{O}^{\nu_1 \nu_2}_i$ is C-even, while ${\cal O}_{\gamma_{\nu_3}}$ is C-odd. 

The next most ``economic'' possibility is to consider a three-point function constructed from the product of $\mathcal{O}^{\mu \nu}_i$ with two fermion bilinear operators ($\mathcal{O}_X$, $\mathcal{O}_Y$) at three distinct spacetime points:
\begin{equation}
G^{\mu \nu}_{i;XY} (x-w, y-w) \equiv \langle \mathcal{O}_i^{\mu \nu} (w) \mathcal{O}_X (x) \mathcal{O}_Y (y) \rangle,
\label{GFs4}
\end{equation}
where the flavor of the fermion (antifermion) field in ${\cal O}_X$ must coincide with the flavor of the antifermion (fermion) field in ${\cal O}_Y$. The above GF depends on two position vectors: $(x-w)$ and $(y-w)$. This fact increases the complexity of the perturbative calculation. This also means that the renormalization factors defined in GIRS may depend on two renormalization 4-vector scales. \textit{A priori}, a possible way of addressing these issues could be to adopt a zero-momentum insertion for one of the three operators. To do this one needs to perform a 4-dimensional integration over the position vector $x$, $y$, or $w$ depending on which operator carries zero momentum. Then, the resulting GF will depend on only one vector. However, such an integration over the whole spacetime causes additional complications: contact terms arise when any two position vectors among $\{x, y, w\}$, coincide, giving additional UV-divergences. To eliminate such divergences, further additive renormalizations would be needed.

One possible alternative way of simplifying the calculation of $G^{\mu \nu}_{i;XY} (x-w, y-w)$, which does not create any contact term is to choose the vector $(y-w)$ to be parallel (or antiparallel) to $(x-w)$ (but $x \neq y$). In this way, $G^{\mu \nu}_{i;XY} $ will depend on a single position vector. A particular example, which we apply in our calculations, is $(y-w) = -(x-w)$ and (without loss of generality) $w=0$:
\begin{equation}
G^{\mu \nu}_{i;XY} (x) \equiv \langle \mathcal{O}_i^{\mu \nu} (0) \mathcal{O}_X (x) O_Y (-x) \rangle.
\label{GFs4b}
\end{equation} 

\begin{figure}[h]
\centering
\epsfig{file=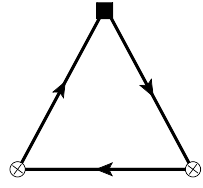,scale=0.45}
\caption{Diagram which contributes to the tree-level Green's functions $G^{\mu \nu}_{2;XY}$. A square denotes insertion of $\mathcal{O}_2^{\mu \nu}$. A cross denotes insertion of a fermion bilinear operator. A diagram having the arrows of the fermion lines in counterclockwise direction must also be considered. 
} \label{EMTtreeXY}
\end{figure}

\begin{figure}[h]
\centering
\epsfig{file=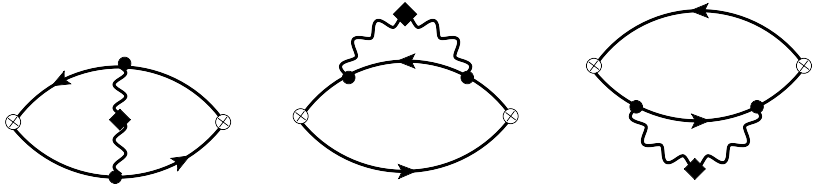,scale=0.65}
\caption{Diagrams which contribute to the one-loop level Green's functions $G^{\mu \nu}_{1;XY}$. A diamond denotes insertion of $\mathcal{O}_1^{\mu \nu}$. A cross denotes insertion of a fermion bilinear operator.
} \label{EMT1XY}
\end{figure}

\begin{figure}[h]
\centering
\epsfig{file=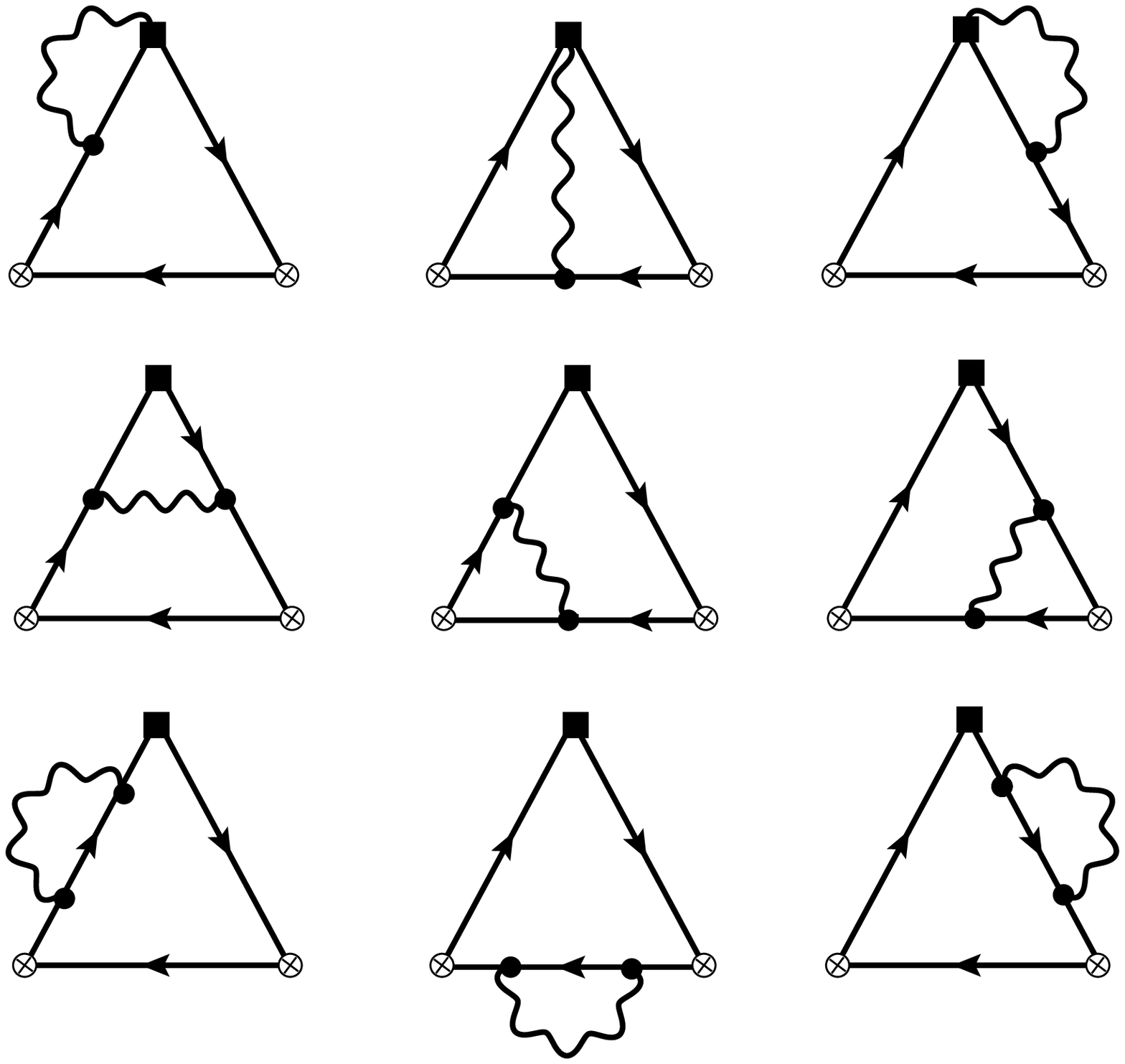,scale=0.58}
\caption{Diagrams which contribute to the one-loop level Green's functions $G^{\mu \nu}_{2;XY}$. A square denotes insertion of $\mathcal{O}_2^{\mu \nu}$. A cross denotes insertion of a fermion bilinear operator. Diagrams having the arrows of the fermion lines in counterclockwise direction must also be considered. 
} \label{EMT2XY}
\end{figure}

The tree-level and one-loop Feynman diagrams contributing to $G^{\mu \nu}_{i;XY}$, $(i=1,2)$, are given in Figs. \ref{EMTtreeXY} and \ref{EMT1XY} - \ref{EMT2XY}, respectively. Note that ${(G^{\mu \nu}_{1;XY})}^{\rm tree} = 0$. A method for calculating the d-dimensional integrals stemming from these Feynman diagrams is described in appendix \ref{Technical aspects of the calculation}.
\clearpage
A most natural set of four conditions for calculating the mixing matrix in GIRS, involving the above GFs, is:
\begin{eqnarray}
&1.& \langle {{\cal O}_1^{\nu_1 \nu_2}}^{\rm GIRS} (x) \ {{\cal O}_1^{\nu_3 \nu_4}}^{\rm GIRS} (y) \rangle \vert_{x-y = \bar{z}} \ = \ 
\langle {{\cal O}_1^{\nu_1 \nu_2}}^{\rm GIRS} (x) \ {{\cal O}_1^{\nu_3 \nu_4}}^{\rm GIRS} (y) \rangle ^{\rm tree} \vert_{x-y = \bar{z}}, \label{cond01} \\
&2.& \langle {{\cal O}_2^{\nu_1 \nu_2}}^{\rm GIRS} (x) \ {{\cal O}_2^{\nu_3 \nu_4}}^{\rm GIRS} (y) \rangle \vert_{x-y = \bar{z}} \ = \ 
\langle {{\cal O}_2^{\nu_1 \nu_2}}^{\rm GIRS} (x) \ {{\cal O}_2^{\nu_3 \nu_4}}^{\rm GIRS} (y) \rangle ^{\rm tree} \vert_{x-y = \bar{z}}, \label{GIRScond2}  \\
&3.& \langle {{\cal O}_1^{\nu_1 \nu_2}}^{\rm GIRS} (x) \ {{\cal O}_2^{\nu_3 \nu_4}}^{\rm GIRS} (y) \rangle \vert_{x-y = \bar{z}} \ = \
\langle {{\cal O}_1^{\nu_1 \nu_2}}^{\rm GIRS} (x) \ {{\cal O}_2^{\nu_3 \nu_4}}^{\rm GIRS} (y) \rangle ^{\rm tree} \vert_{x-y = \bar{z}} \ = \ 0,   \\
&4.& \langle {{\cal O}_1^{\nu_1 \nu_2}}^{\rm GIRS} (0) \ {\cal O}^{\rm GIRS}_X (x) \ {{\cal O}^{\rm GIRS}_Y} (-x) \rangle \vert_{x = \bar{z}} \ = \ 
\langle {{\cal O}_1^{\nu_1 \nu_2}}^{\rm GIRS} (0) \ {\cal O}^{\rm GIRS}_X (x) \ {{\cal O}^{\rm GIRS}_Y} (-x) \rangle^{\rm tree} \vert_{x = \bar{z}} \ = \ 0, \label{cond04}   
\end{eqnarray}
where $\nu_1 \neq \nu_2$ and $\nu_3 \neq \nu_4$. Alternatively, we can replace the second condition (Eq.~\eqref{GIRScond2}) with:
\be
\langle {{\cal O}_2^{\nu_1 \nu_2}}^{\rm GIRS} (0) \ {\cal O}^{\rm GIRS}_X (x) \ {{\cal O}^{\rm GIRS}_Y} (-x) \rangle \vert_{x = \bar{z}} \ = \ 
\langle {{\cal O}_2^{\nu_1 \nu_2}}^{\rm GIRS} (0) \ {\cal O}^{\rm GIRS}_X (x) \ {{\cal O}^{\rm GIRS}_Y} (-x) \rangle^{\rm tree} \vert_{x = \bar{z}}.
\ee
We only need to make one convenient choice for X and Y. All other choices should be related by conversion factors; it is useful to check that these factors are indeed finite. Note that some choices of X and Y may give vanishing contributions, depending on the transformation properties of X and Y under rotations, parity and charge conjugation. The conditions~\eqref{cond01} -~\eqref{cond04} can be written in the following explicit form\footnote{For simplicity, we omit superscripts referring to the regularization and renormalization scheme, as well as Lorentz indices. We also omit the dependence on spacetime coordinates.}:
\begin{eqnarray}
Z_{11}^2 \ G_{11} + 2 \ Z_{11} \ Z_{12} \ G_{12} + Z_{12}^2 \ G_{22} &=& G_{11}^{\rm tree}, \label{cond1}\\
Z_{21}^2 \ G_{11} + 2 \ Z_{21} \ Z_{22} \ G_{12} + Z_{22}^2 \ G_{22} &=& G_{22}^{\rm tree}, \\
Z_{11} \ Z_{21} \ G_{11} + (Z_{11} \ Z_{22} + Z_{12} \ Z_{21}) \ G_{12} + Z_{12} \ Z_{22} \ G_{22} &=& G_{12}^{\rm tree} = 0, \\
Z_X Z_Y \ (Z_{11} \ G_{1; X Y} + Z_{12} \ G_{2; X Y}) &=& G_{1; X Y}^{\rm tree} = 0. \label{cond4}  
\end{eqnarray}
The four elements of the mixing matrix $Z_{ij}$ can be obtained by solving the above system of 4 equations, once all GFs on the left-hand sides have been determined via numerical simulations. Note that renormalization factors $Z_X$ and $Z_Y$, appearing in Eq.~\eqref{cond4}, are eliminated and thus they do not contribute to the calculation of $Z_{ij}$.

\smallskip

\vspace{0.1cm}A proper extension of t-GIRS, analogous to what was defined for the renormalization of fermion bilinears (see Eq. \eqref{RC2}), can be applied in this case, leading to the following conditions:
\begin{eqnarray}
&1.& \int d^3\vec{x} \ \langle {{\cal O}_1^{\nu_1 \nu_2}}^{\rm t-GIRS} (\vec{x},0) \ {{\cal O}_1^{\nu_1 \nu_2}}^{\rm t-GIRS} (\vec{0},\bar{t}) \rangle \ = \
\int d^3\vec{x} \ \langle {{\cal O}_1^{\nu_1 \nu_2}}^{\rm t-GIRS} (\vec{x},0) \ {{\cal O}_1^{\nu_1 \nu_2}}^{\rm t-GIRS} (\vec{0},\bar{t}) \rangle ^{\rm tree}, \\
&2.& \int d^3\vec{x} \ \langle {{\cal O}_2^{\nu_1 \nu_2}}^{\rm t-GIRS} (\vec{x},0) \ {{\cal O}_2^{\nu_1 \nu_2}}^{\rm t-GIRS} (\vec{0},\bar{t}) \rangle \ = \ 
\int d^3\vec{x} \ \langle {{\cal O}_2^{\nu_1 \nu_2}}^{\rm t-GIRS} (\vec{x},0) \ {{\cal O}_2^{\nu_1 \nu_2}}^{\rm t-GIRS} (\vec{0},\bar{t}) \rangle ^{\rm tree},   \\
&3.& \int d^3\vec{x} \ \langle {{\cal O}_1^{\nu_1 \nu_2}}^{\rm t-GIRS} (\vec{x},0) \ {{\cal O}_2^{\nu_1 \nu_2}}^{\rm t-GIRS} (\vec{0},\bar{t}) \rangle \ = \
\int d^3\vec{x} \ \langle {{\cal O}_1^{\nu_1 \nu_2}}^{\rm t-GIRS} (\vec{x},0) \ {{\cal O}_2^{\nu_1 \nu_2}}^{\rm t-GIRS} (\vec{0},\bar{t}) \rangle ^{\rm tree} = 0,   \\
&4.& \int d^3\vec{x} \ \langle {{\cal O}_1^{\nu_1 \nu_2}}^{\rm t-GIRS} (\vec{0},0) \ {\cal O}^{\rm t-GIRS}_X (\vec{x},\bar{t}) \ {{\cal O}^{\rm t-GIRS}_Y} (-\vec{x},-\bar{t}) \rangle \ = \nonumber \\
&& \int d^3\vec{x} \ \langle {{\cal O}_1^{\nu_1 \nu_2}}^{\rm t-GIRS} (\vec{0},0) \ {\cal O}^{\rm t-GIRS}_X (\vec{x},\bar{t}) \ {{\cal O}^{\rm t-GIRS}_Y} (-\vec{x},-\bar{t}) \rangle^{\rm tree} \ = \ 0. \label{tGIRScond4}
\end{eqnarray}
No summation over $\nu_1$, $\nu_2$ is implied. Depending on the choice of $X$ and $Y$, the fourth condition of t-GIRS may involve odd integrals, which give zero. For example, using $X = Y = \openone$ will lead to two structures: (i) $\delta_{\nu_1 \nu_2}$, which vanishes since we study the nondiagonal elements ($\nu_1 \neq \nu_2$) of $\mathcal{O}_i^{\nu_1 \nu_2}$, and (ii) $x_{\nu_1} x_{\nu_2}$, which will vanish upon integration over $\vec{x}$ for any choices of $\nu_1$, $\nu_2$. In such cases, an appropriate variant of the fourth condition can be applied; e.g., for the case $X = Y = \openone$, a possible alternative condition, in place of Eq.~\eqref{tGIRScond4}, is:
\bea
&& \int d^3\vec{x} \ \frac{x_{\nu_1} x_{\nu_2}}{x^2} \ \langle {{\cal O}_1^{\nu_1 \nu_2}}^{\rm t-GIRS} (\vec{0},0) \ {\cal O}^{\rm t-GIRS}_{\openone} (\vec{x},\bar{t}) \ {{\cal O}^{\rm t-GIRS}_{\openone}} (-\vec{x},-\bar{t}) \rangle \ = \nonumber \\ 
&& \int d^3\vec{x} \ \frac{x_{\nu_1} x_{\nu_2}}{x^2} \ \langle {{\cal O}_1^{\nu_1 \nu_2}}^{\rm t-GIRS} (\vec{0},0) \ {\cal O}^{\rm t-GIRS}_{\openone} (\vec{x},\bar{t}) \ {{\cal O}^{\rm t-GIRS}_{\openone}} (-\vec{x},-\bar{t}) \rangle^{\rm tree} \ = \ 0,
\label{tGIRScond4_2}
\eea
(no summation over $\nu_1$, $\nu_2$ is implied), or
\bea
&& \int d^3\vec{x} \ e^{i \vec{p} \cdot \vec{x}} \ \langle {{\cal O}_1^{\nu_1 \nu_2}}^{\rm t-GIRS} (\vec{0},0) \ {\cal O}^{\rm t-GIRS}_{\openone} (\vec{x},\bar{t}) \ {{\cal O}^{\rm t-GIRS}_{\openone}} (-\vec{x},-\bar{t}) \rangle \ = \nonumber \\
&& \int d^3\vec{x} \ e^{i \vec{p} \cdot \vec{x}} \ \langle {{\cal O}_1^{\nu_1 \nu_2}}^{\rm t-GIRS} (\vec{0},0) \ {\cal O}^{\rm t-GIRS}_{\openone} (\vec{x},\bar{t}) \ {{\cal O}^{\rm t-GIRS}_{\openone}} (-\vec{x},-\bar{t}) \rangle^{\rm tree} \ = \ 0,
\label{tGIRScond4_3}
\eea
for a fixed choice of the 3-vector $\vec{p}$. These variant schemes lead to relations analogous to Eqs. (\ref{cond1} - \ref{cond4}) for the determination of $Z_{ij}$. In this work, we present one-loop results for three-point GFs with ($X$, $Y$) $=$ ($\openone$, $\openone$), ($\gamma_5$, $\gamma_5$), ($\gamma_{\nu_3}$, $\gamma_{\nu_4}$), ($\gamma_5 \gamma_{\nu_3}$, $\gamma_5 \gamma_{\nu_4}$), before performing integration over $\vec{x}$. It is straightforward to integrate all these results over $\vec{x}$; we do so for some specific cases (see next section), which are likely the most appropriate for nonperturbative investigations.

After calculating the mixing matrix, we extract the conversion factors between the GIRS (or t-GIRS) and the $\MSbar$ scheme for the EMT operators; they have a $2 \times 2$ matrix form:
\be
\begin{pmatrix}
{\mathcal{O}_1^{\mu \nu}}^\MSbar \\
\\
{\mathcal{O}_2^{\mu \nu}}^\MSbar
\end{pmatrix} = 
\begin{pmatrix}
C_{11}^{{\rm GIRS},\MSbar} & C_{12}^{{\rm GIRS},\MSbar} \\
\\
C_{21}^{{\rm GIRS},\MSbar} & C_{22}^{{\rm GIRS},\MSbar}
\end{pmatrix} \cdot
\begin{pmatrix}
{\mathcal{O}_1^{\mu \nu}}^{\rm GIRS} \\
\\
{\mathcal{O}_2^{\mu \nu}}^{\rm GIRS}
\end{pmatrix} \ \Rightarrow
\ee
\be
\begin{pmatrix}
C_{11}^{{\rm GIRS},\MSbar} & C_{12}^{{\rm GIRS},\MSbar} \\
\\
C_{21}^{{\rm GIRS},\MSbar} & C_{22}^{{\rm GIRS},\MSbar}
\end{pmatrix} = \begin{pmatrix}
Z_{11}^{B,\MSbar} & Z_{12}^{B,\MSbar} \\
\\
Z_{21}^{B,\MSbar} & Z_{22}^{B,\MSbar}
\end{pmatrix} \cdot \begin{pmatrix}
Z_{11}^{B,{\rm GIRS}} & Z_{12}^{B,{\rm GIRS}} \\
\\
Z_{21}^{B,{\rm GIRS}} & Z_{22}^{B,{\rm GIRS}}
\end{pmatrix}^{-1}.
\label{convEMT}
\ee

The Z factors in Eq.~\eqref{convEMT} can be computed in any regularization ``$B$''; given that we are making contact with $\MSbar$, the most natural choice for $B$ is dimensional regularization.

\subsection{Results}

In this subsection, we present our results (up to one loop) for the bare GFs $\langle\mathcal{O}^{\nu_1 \nu_2}_i(x)\mathcal{O}^{\nu_3 \nu_4}_j(y)\rangle$, and $\langle \mathcal{O}^{\nu_1 \nu_2}_i (0) \mathcal{O}_X (x) \mathcal{O}_Y (-x) \rangle$ for $i, j = 1,2$ and $(X, Y) =$ ($\openone$, $\openone$), ($\gamma_5$, $\gamma_5$), ($\gamma_{\nu_3}$, $\gamma_{\nu_4}$), ($\gamma_5 \gamma_{\nu_3}$, $\gamma_5 \gamma_{\nu_4}$), as well as the conversion factors between all variants of GIRS and $\MSbar$. The results are expressed in terms of the following Lorentz structures :
\bea
s^{[1]}_{\nu_1 \nu_2} (x) \ \ \ \ \, &\equiv & \frac{x_{\nu_1} x_{\nu_2}}{x^2}, \\
s^{[2]}_{\nu_1 \nu_2 \nu_3 \nu_4} (x) &\equiv &  \frac{x_{\nu_1} x_{\nu_2}}{x^2} \ \delta_{\nu_3 \nu_4}, \\
s^{[3]}_{\nu_1 \nu_2 \nu_3 \nu_4} \ \ \ \ &\equiv & \delta_{\nu_1 \nu_3} \delta_{\nu_2 \nu_4} + \delta_{\nu_1 \nu_4} \delta_{\nu_2 \nu_3}, \\
s^{[4]}_{\nu_1 \nu_2 \nu_3 \nu_4} (x) &\equiv & \delta_{\nu_1 \nu_3} \frac{x_{\nu_2} x_{\nu_4}}{x^2} + \delta_{\nu_1 \nu_4} \frac{x_{\nu_2} x_{\nu_3}}{x^2}  + \delta_{\nu_2 \nu_3} \frac{x_{\nu_1} x_{\nu_4}}{x^2} + \delta_{\nu_2 \nu_4} \frac{x_{\nu_1} x_{\nu_3}}{x^2}, \\
s^{[5]}_{\nu_1 \nu_2 \nu_3 \nu_4} (x) &\equiv & \frac{x_{\nu_1} x_{\nu_2} x_{\nu_3} x_{\nu_4}}{{(x^2)}^2}.
\eea
The resulting expressions\footnote{For brevity, decimal numbers in our results are presented only with six digits after the decimal point; they are known to higher accuracy.} for the bare GFs are ($z \equiv y-x$):
\bea
\langle\mathcal{O}^{\nu_1 \nu_2}_1(x)\mathcal{O}^{\nu_3 \nu_4}_1(y)\rangle &=& \Big(s^{[3]}_{\nu_1 \nu_2 \nu_3 \nu_4} - 2 \ s^{[4]}_{\nu_1 \nu_2 \nu_3 \nu_4} (z) + 8 \ s^{[5]}_{\nu_1 \nu_2 \nu_3 \nu_4} (z)\Big) \ \frac{4 N_c \ C_F \ (1-2\varepsilon) \ {(\Gamma (2-\varepsilon))}^2}{\pi^{4 - 2 \varepsilon} {(z^2)}^{4 - 2 \varepsilon}} \times \nonumber \\ 
&& \ \Bigg[1 - \frac{g_\MSbar^2}{16\pi^2} \Big(\frac{4 N_f}{3} \Big(\frac{1}{ \varepsilon} + 2 \gamma_E - 2 \ln (2) + \ln (\bar{\mu}^2 z^2) + \frac{1}{6} \Big) + \frac{20 N_c}{9} \Big) \Bigg] + \nonumber \\
&& \quad \mathcal{O} (\varepsilon^2, g_\MSbar^0) \ + \ \mathcal{O} (\varepsilon^1, g_\MSbar^2) \ + \ \mathcal{O} (g_\MSbar^4),
\label{O1O1}
\eea
\bea
\langle\mathcal{O}^{\nu_1 \nu_2}_1(x)\mathcal{O}^{\nu_3 \nu_4}_2(y)\rangle &=& \Big(s^{[3]}_{\nu_1 \nu_2 \nu_3 \nu_4} - 2 \ s^{[4]}_{\nu_1 \nu_2 \nu_3 \nu_4} (z) + 8 \ s^{[5]}_{\nu_1 \nu_2 \nu_3 \nu_4} (z)\Big) \ \frac{N_c \ N_f \ (2-\varepsilon) \ {(\Gamma (2-\varepsilon))}^2 \ (1- \frac{3}{4} \varepsilon)}{2 \ \pi^{4 - 2 \varepsilon} {(z^2)}^{4 - 2 \varepsilon}} \times \nonumber \\
&& \ \ \frac{g_\MSbar^2}{16\pi^2} \ \frac{16 C_F}{3} \left(\frac{1}{\varepsilon} + 2 \gamma_E - 2 \ln (2) + \ln (\bar{\mu}^2 z^2) - \frac{1}{6} \right) + \ \mathcal{O} (\varepsilon^1, g_\MSbar^2) \ + \ \mathcal{O} (g_\MSbar^4),
\label{O1O2}
\eea
\bea
\langle\mathcal{O}^{\nu_1 \nu_2}_2(x)\mathcal{O}^{\nu_3 \nu_4}_2(y)\rangle &=& \Big(s^{[3]}_{\nu_1 \nu_2 \nu_3 \nu_4} - 2 \ s^{[4]}_{\nu_1 \nu_2 \nu_3 \nu_4} (z) + 8 \ s^{[5]}_{\nu_1 \nu_2 \nu_3 \nu_4} (z)\Big) \ \frac{N_c \ N_f \ (2-\varepsilon) \ {(\Gamma (2-\varepsilon))}^2}{2 \ \pi^{4 - 2 \varepsilon} {(z^2)}^{4 - 2 \varepsilon}} \times \\
&& \ \Bigg[1 - \frac{g_\MSbar^2}{16\pi^2} \left(\frac{16 C_F}{3}\right) \Big(\frac{1}{ \varepsilon} + 2 \gamma_E - 2 \ln (2) + \ln (\bar{\mu}^2 z^2) - \frac{59}{48} \Big) \Bigg] + \nonumber \\
&& \quad \mathcal{O} (\varepsilon^2, g_\MSbar^0) \ + \ \mathcal{O} (\varepsilon^1, g_\MSbar^2) \ + \ \mathcal{O} (g_\MSbar^4),
\label{O2O2}
\eea
\bea
\langle \mathcal{O}^{\nu_1 \nu_2}_1 (0) \mathcal{O}_{\openone} (x) \mathcal{O}_{\openone} (-x) \rangle &=& - s^{[1]}_{\nu_1 \nu_2} (x) \ \frac{N_c \ N_f \ {(\Gamma (2-\varepsilon))}^2 \ \Gamma (3-\varepsilon)}{2^{2 - 2 \varepsilon} \ \pi^{6 - 3 \varepsilon} \ {(x^2)}^{5 - 3 \varepsilon}} \ \frac{g_\MSbar^2}{16\pi^2} \ \frac{8 C_F}{3} \left[\frac{1}{\varepsilon} + \ln (\bar{\mu}^2 x^2) -0.701491 \right] + \nonumber \\
&& \ \ \mathcal{O} (\varepsilon^1, g_\MSbar^2) \ + \ \mathcal{O} (g_\MSbar^4),
\eea
\be
\langle \mathcal{O}^{\nu_1 \nu_2}_1 (0) \mathcal{O}_{\gamma_5} (x) \mathcal{O}_{\gamma_5} (-x) \rangle = - \langle \mathcal{O}^{\nu_1 \nu_2}_1 (0) \mathcal{O}_{\openone} (x) \mathcal{O}_{\openone} (-x) \rangle,
\ee
\bea
\langle \mathcal{O}^{\nu_1 \nu_2}_1 (0) \mathcal{O}_{\gamma_{\nu_3}} (x) \mathcal{O}_{\gamma_{\nu_4}} (-x) \rangle &=& \frac{N_c \ N_f \ {(\Gamma (2-\varepsilon))}^2 \ \Gamma (3-\varepsilon)}{2^{3 - 2 \varepsilon} \ \pi^{6 - 3 \varepsilon} \ {(x^2)}^{5 - 3 \varepsilon}} \ \frac{g_\MSbar^2}{16\pi^2} \ \frac{8 C_F}{3} \times \nonumber \\
&&  \Bigg[ \Big(s^{[4]}_{\nu_1 \nu_2 \nu_3 \nu_4} (x) + 2 \ s^{[2]}_{\nu_1 \nu_2 \nu_3 \nu_4} (x) - 8 \ s^{[5]}_{\nu_1 \nu_2 \nu_3 \nu_4} (x) \Big) \ \Big(\frac{1}{\varepsilon} + \ln (\bar{\mu}^2 x^2) - 1.701491 \Big) \nonumber \\
&& \quad + \frac{1}{2} \ s^{[2]}_{\nu_1 \nu_2 \nu_3 \nu_4} (x) + \frac{3}{4} \ s^{[3]}_{\nu_1 \nu_2 \nu_3 \nu_4} - s^{[4]}_{\nu_1 \nu_2 \nu_3 \nu_4} (x) \Bigg] + \ \mathcal{O} (\varepsilon^1, g_\MSbar^2) \ + \ \mathcal{O} (g_\MSbar^4),
\eea
\be
\langle \mathcal{O}^{\nu_1 \nu_2}_1 (0) \mathcal{O}_{\gamma_5 \gamma_{\nu_3}} (x) \mathcal{O}_{\gamma_5 \gamma_{\nu_4}} (-x) \rangle = \langle \mathcal{O}^{\nu_1 \nu_2}_1 (0) \mathcal{O}_{\gamma_{\nu_3}} (x) \mathcal{O}_{\gamma_{\nu_4}} (-x) \rangle,
\ee
\bea
\langle \mathcal{O}^{\nu_1 \nu_2}_2 (0) \mathcal{O}_{\openone} (x) \mathcal{O}_{\openone} (-x) \rangle &=& - s^{[1]}_{\nu_1 \nu_2} (x) \ \frac{N_c \ N_f \ {(\Gamma (2-\varepsilon))}^2 \ \Gamma (3-\varepsilon)}{2^{2 - 2 \varepsilon} \ \pi^{6 - 3 \varepsilon} \ {(x^2)}^{5 - 3 \varepsilon}} \times \nonumber \\
&& \ \Bigg[ 1 + \frac{g_\MSbar^2}{16\pi^2}  \frac{10 C_F}{3} \left(\frac{1}{\varepsilon} + \ln (\bar{\mu}^2 x^2) + 2.639169 \right) \Bigg] + \nonumber \\
&& \quad \mathcal{O} (\varepsilon^1, g_\MSbar^2) \ + \ \mathcal{O} (g_\MSbar^4),
\eea
\bea
\langle \mathcal{O}^{\nu_1 \nu_2}_2 (0) \mathcal{O}_{\gamma_5} (x) \mathcal{O}_{\gamma_5} (-x) \rangle &=& -  \langle \mathcal{O}^{\nu_1 \nu_2}_2 (0) \mathcal{O}_{\openone} (x) \mathcal{O}_{\openone} (-x) \rangle \nonumber \\
&& + \ {\rm hv} \ s^{[1]}_{\nu_1 \nu_2} (x) \ \frac{N_c \ N_f \ {(\Gamma (2-\varepsilon))}^2 \ \Gamma (3-\varepsilon)}{2^{2 - 2 \varepsilon} \ \pi^{6 - 3 \varepsilon} {(x^2)}^{5 - 3 \varepsilon}} \ \frac{g_\MSbar^2}{16\pi^2} \ 16 \ C_F \ + \ \mathcal{O} (g_\MSbar^4),
\eea
\bea
\langle \mathcal{O}^{\nu_1 \nu_2}_2 (0) \mathcal{O}_{\gamma_{\nu_3}} (x) \mathcal{O}_{\gamma_{\nu_4}} (-x) \rangle &=& \frac{N_c \ N_f \ {(\Gamma (2-\varepsilon))}^2 \ \Gamma (3-\varepsilon)}{2^{3 - 2 \varepsilon} \ \pi^{6 - 3 \varepsilon} {(x^2)}^{5 - 3 \varepsilon}} \times \nonumber \\
&&  \Bigg[ \Big(s^{[4]}_{\nu_1 \nu_2 \nu_3 \nu_4} (x) + 2 \ s^{[2]}_{\nu_1 \nu_2 \nu_3 \nu_4} (x) - 8 \ s^{[5]}_{\nu_1 \nu_2 \nu_3 \nu_4} (x) \Big) \times \nonumber \\
&& \quad  \Bigg(1 - \frac{g_\MSbar^2}{16\pi^2} \ \frac{8 C_F}{3} \ \Big(\frac{1}{\varepsilon} + \ln (\bar{\mu}^2 x^2) - 3.201491 \Big) \Bigg) \nonumber \\
&& \quad - \frac{g_\MSbar^2}{16\pi^2} \ \frac{8 C_F}{3} \ \Big(\frac{11}{4} \ s^{[2]}_{\nu_1 \nu_2 \nu_3 \nu_4} (x) + \frac{9}{8} \ s^{[3]}_{\nu_1 \nu_2 \nu_3 \nu_4} - s^{[4]}_{\nu_1 \nu_2 \nu_3 \nu_4} (x) \Bigg) \Bigg] + \nonumber \\
&& \quad \mathcal{O} (\varepsilon^1, g_\MSbar^2) \ + \ \mathcal{O} (g_\MSbar^4),
\eea
\bea
\langle \mathcal{O}^{\nu_1 \nu_2}_2 (0) \mathcal{O}_{\gamma_5 \gamma_{\nu_3}} (x) \mathcal{O}_{\gamma_5 \gamma_{\nu_4}} (-x) \rangle &=& \langle \mathcal{O}^{\nu_1 \nu_2}_2 (0) \mathcal{O}_{\gamma_{\nu_3}} (x) \mathcal{O}_{\gamma_{\nu_4}} (-x) \rangle + \ {\rm hv} \ \frac{N_c \ N_f \ {(\Gamma (2-\varepsilon))}^2 \ \Gamma (3-\varepsilon)}{2^{3 - 2 \varepsilon} \ \pi^{6 - 3 \varepsilon} {(x^2)}^{5 - 3 \varepsilon}} \times \nonumber \\
&&  \Big(s^{[4]}_{\nu_1 \nu_2 \nu_3 \nu_4} (x) + 2 \ s^{[2]}_{\nu_1 \nu_2 \nu_3 \nu_4} (x) - 8 \ s^{[5]}_{\nu_1 \nu_2 \nu_3 \nu_4} (x) \Big) \frac{g_\MSbar^2}{16\pi^2} \ 8 \ C_F \ + \ \mathcal{O} (g_\MSbar^4).
\label{O2AA}
\eea

The above GFs lead to the following results for $Z_{ij}^{{\rm DR}, \MSbar}$:
\bea
Z_{11}^{{\rm DR},\MSbar} &=& 1 \ + \ \frac{g^2}{16 \pi^2} \ \frac{1}{\varepsilon} \ \left(\phantom{+} \frac{2}{3} N_f\right) \ + \ \mathcal{O} (g_\MSbar^4), \\
Z_{12}^{{\rm DR},\MSbar} &=&  \ \ \ \ \ \ \ \frac{g^2}{16 \pi^2} \ \frac{1}{\varepsilon} \ \left(- \frac{8}{3} C_F\right) \ + \ \mathcal{O} (g_\MSbar^4), \\
Z_{21}^{{\rm DR},\MSbar} &=& \ \ \ \ \ \ \ \frac{g^2}{16 \pi^2} \ \frac{1}{\varepsilon} \ \left(- \frac{2}{3} N_f\right) \ + \ \mathcal{O} (g_\MSbar^4), \\
Z_{22}^{{\rm DR},\MSbar} &=& 1 \ + \ \frac{g^2}{16 \pi^2} \ \frac{1}{\varepsilon} \ \left(\phantom{+} \frac{8}{3} C_F\right) \ + \ \mathcal{O} (g_\MSbar^4).
\eea
These are consistent with the results found using GFs with elementary external fields~\cite{Panagopoulos:2020qcn}. The corresponding $\MSbar$-renormalized GFs can be obtained by removing $1/\varepsilon$ terms from Eqs.~\eqref{O1O1} -~\eqref{O2AA} and by taking the na{\"i}ve limit $\varepsilon \rightarrow 0$ in the remaining terms.

By solving the system of four equations (\ref{cond1} - \ref{cond4}), we extract the conversion factors between different variants of GIRS and $\MSbar$. Below, we present results for five specific variants of GIRS. All these variants are expected to lead to the same $\MSbar$-renormalized operators, but their respective numerical signals may favor one variant over the others.
\begin{enumerate}
\item $\underline{{\rm GIRS}_1}$: $X = Y = \openone$; no integration over $\vec{x}$. \\
Similarly, by choosing $X = Y = \gamma_5$, we will arrive at the same one-loop conversion factors, since the ``hv'' coefficient, which would have made a difference, appears only in the one-loop GF $\langle \mathcal{O}_2^{\nu_1 \nu_2} (0) \mathcal{O}_{\gamma_5} (x) \mathcal{O}_{\gamma_5} (-x) \rangle$, which does not contribute in the calculation of the conversion factors to one loop. Nevertheless, numerical data will be much different; thus, this provides for an interesting comparison of the corresponding $\MSbar$-renormalized GFs, as gotten from the lattice.
\item $\underline{{\rm GIRS}_2}$: $X = \gamma_{\nu_3}, Y = \gamma_{\nu_4}$; no integration over $\vec{x}$; $\nu_1, \nu_2, \nu_3, \nu_4$ are all different. \\
Since the three-point functions $\langle \mathcal{O}_i^{\nu_1 \nu_2} (0) \mathcal{O}_{\gamma_{\nu_3}} (x) \mathcal{O}_{\gamma_{\nu_4}} (-x) \rangle$ have more than one Lorentz structures, it is necessary to isolate a structure by using projectors or by making specific choices for the indices $\nu_1$ -- $\nu_4$ and/or the components of $x$. In this variant, we isolate the structure $s^{[5]}_{\nu_1 \nu_2 \nu_3 \nu_4} (x)$ by choosing $\nu_1, \nu_2, \nu_3, \nu_4$ to be all different. In this case all four components of $x$ must be nonzero. Similarly, the choice of $X = \gamma_5 \gamma_{\nu_3}, Y = \gamma_5 \gamma_{\nu_4}$ will give the same conversion factors to one loop.
\item $\underline{{\rm GIRS}_3}$: $X = \gamma_{\nu_3}, Y = \gamma_{\nu_4}$; no integration over $\vec{x}$; \ $\nu_3 = \nu_4, \nu_3 \neq (\nu_1, \nu_2)$; and $x_{\nu_3} = 0$. \\
This variant is similar to ${\rm GIRS}_2$ with the difference of isolating the structure $s^{[2]}_{\nu_1 \nu_2 \nu_3 \nu_4} (x)$.
\item $\underline{{\rm t-GIRS}_1}$: $X = \gamma_{\nu_3}, Y = \gamma_{\nu_4}$; integration over $\vec{x}$; $\nu_1 = \nu_3, \nu_2 = \nu_4$; $\nu_1, \nu_2$ are both spatial. \\
The integration over the spatial components of $x$ will give zero unless the four indices are paired (both in the two-point and three-point GFs), e.g., $\nu_1 = \nu_3, \nu_2 = \nu_4$. In principle, there are two distinct possibilities: $\nu_1, \nu_2$ are both spatial or $\nu_1$ is spatial and $\nu_2$ is temporal. However, the latter case will be impossible to satisfy, since the combination $\Big(s^{[3]}_{\nu_1 \nu_2 \nu_3 \nu_4} - 2 \ s^{[4]}_{\nu_1 \nu_2 \nu_3 \nu_4} (z) + 8 \ s^{[5]}_{\nu_1 \nu_2 \nu_3 \nu_4} (z)\Big) / {(z^2)}^4$ appearing in the two-point functions [see Eqs. (~\ref{O1O1} -~\ref{O2O2})] vanishes upon integration over spatial components.
\item $\underline{{\rm t-GIRS}_2}$: $X = Y = \openone$; integration over $\vec{x}$; $\nu_1 = \nu_3, \nu_2 = \nu_4$; $\nu_1, \nu_2$ are both spatial; projector: $x_{\nu_1} x_{\nu_2} / x^2$ (see Eq.~\eqref{tGIRScond4_2}).
\end{enumerate}
The conversion factors for the above variants of GIRS are given below:
\bea
C_{11}^{{\rm GIRS}_i, \MSbar} &=& 1 - \frac{g_\MSbar^2}{16 \pi^2} \Big[ \frac{10}{9} N_c + c_{11} N_f + \frac{2}{3} N_f \ln (\bar{\mu}^2 \bar{z}^2) \Big] + \mathcal{O} (g_\MSbar^4), \label{C11GIRS1}\\
C_{12}^{{\rm GIRS}_i, \MSbar} &=& \ \ - \frac{g_\MSbar^2}{16 \pi^2} C_F \Big[c_{12} - \frac{8}{3} \ln (\bar{\mu}^2 \bar{z}^2) \Big] + \mathcal{O} (g_\MSbar^4), \\
C_{21}^{{\rm GIRS}_i, \MSbar} &=& \ \ - \frac{g_\MSbar^2}{16 \pi^2} N_f \Big[c_{21} - \frac{2}{3} \ln (\bar{\mu}^2 \bar{z}^2) \Big] + \mathcal{O} (g_\MSbar^4), \\
C_{22}^{{\rm GIRS}_i, \MSbar} &=& 1 - \frac{g_\MSbar^2}{16 \pi^2} C_F \Big[c_{22} + \frac{8}{3} \ln (\bar{\mu}^2 \bar{z}^2) \Big] + \mathcal{O} (g_\MSbar^4), \\
C_{11}^{{\rm t-GIRS}_j, \MSbar} &=& 1 - \frac{g_\MSbar^2}{16 \pi^2} \Big[ \frac{10}{9} N_c + c_{11} N_f + \frac{2}{3} N_f \ln (\bar{\mu}^2 \bar{t}^2) \Big] + \mathcal{O} (g_\MSbar^4), \\
C_{12}^{{\rm t-GIRS}_j, \MSbar} &=& \ \ - \frac{g_\MSbar^2}{16 \pi^2} C_F \Big[c_{12} - \frac{8}{3} \ln (\bar{\mu}^2 \bar{t}^2) \Big] + \mathcal{O} (g_\MSbar^4), \\
C_{21}^{{\rm t-GIRS}_j, \MSbar} &=& \ \ - \frac{g_\MSbar^2}{16 \pi^2} N_f \Big[c_{21} - \frac{2}{3} \ln (\bar{\mu}^2 \bar{t}^2) \Big] + \mathcal{O} (g_\MSbar^4), \\
C_{22}^{{\rm t-GIRS}_j, \MSbar} &=& 1 - \frac{g_\MSbar^2}{16 \pi^2} C_F \Big[c_{22} + \frac{8}{3} \ln (\bar{\mu}^2 \bar{t}^2) \Big] + \mathcal{O} (g_\MSbar^4),
\label{C22tGIRS2}
\eea
where $i= 1,2,3$, $j=1,2$ and coefficients $c_{kl}$ are given in Table~\ref{Tab:cij} for each variant of GIRS.

\begin{table}[H]
\centering
\begin{tabular}{M{1.5cm} || M{2cm} || M{2cm} || M{2cm} || M{2cm} || M{2cm} N} 
\hline
\hline
& & & & & \\
 & ${\rm GIRS}_1$ & ${\rm GIRS}_2$ & ${\rm GIRS}_3$ & ${\rm t-GIRS}_1$ & ${\rm t-GIRS}_2$ \\[5pt] \hline
 & & & & & \\
$c_{11}$ &  -0.043464 &  -0.043464 &  -0.043464 & \ 0.236288 & \ 0.236288 \\[15pt] \hline
& & & & & \\
$c_{12}$ & \ 1.870642 & \ 4.537309 & \ 3.870642 &  -7.848365 &  -0.181699 \\[15pt] \hline
& & & & & \\
$c_{21}$ & \ 0.063712 &  -0.602954 &  -0.436288 & \ 1.933961 & \ 0.017294 \\[15pt] \hline
& & & & & \\
$c_{22}$ &  -3.896079 &  -3.896079 &  -3.896079 &  -2.777072 &  -2.777072  \\[15pt] \hline
\hline
\end{tabular}
 \caption{Values of the one-loop coefficient $c_{ij}$ in the definition of the conversion factors of EMT operators, given in Eqs. (~\ref{C11GIRS1} -~\ref{C22tGIRS2}).} \label{Tab:cij}
\end{table}

Use of Eq.~\eqref{tGIRScond4_3} as an alternative renormalization condition requires the integration of various expressions of the form:
\be
\frac{e^{i \vec{p} \cdot \vec{x}} \ x_{\mu_1} \ x_{\mu_2} \ldots \ x_{\mu_j} \ {(\ln(x^2))}^i}{{(x^2)}^k}, \qquad (i,j,k: {\rm nonnegative \ integers})
\ee
over spatial components of $x=(\vec{x},t)$. All these integrals can be performed by using the following generating integral function:
\be
\int d^3 x \frac{e^{i \vec{p} \cdot \vec{x}}}{{(x^2)}^a} = \frac{2^{5/2 - a} \pi^{3/2}}{\Gamma (a)} {\left(|p| / |t|\right)}^{-3/2 + a} K_{3/2 - a} \left(|p| |t|\right),
\ee
where $K_\nu (z)$ is the modified Bessel function of the second kind and $a$ may take noninteger values. Then, differentiating with respect to $a$ or to individual components of $p$, we can calculate all necessary integrals arising in t-GIRS.

\section{Summary}
\label{summary}
In this paper, we study a gauge-invariant, mass-independent renormalization scheme (GIRS) for composite operators, which is applicable in both perturbative and nonperturbative studies. This is an extended version of the coordinate space (X-space) renormalization scheme studied in, e.g., Refs.~\cite{Jansen:1995ck, Gimenez:2004me}. This scheme involves vacuum expectation values of products of gauge-invariant operators located at different spacetime points. The expectation values are gauge-independent and thus, gauge fixing is not needed in this scheme. Also, gauge-variant operators, which may mix with gauge-invariant operators, do not contribute in such Green's functions; as a consequence, they can be safely excluded, leading to a reduced set of mixing operators. In this work, we apply GIRS in the renormalization of fermion bilinear operators, as well as in the renormalization and mixing of the gluon and quark parts of the QCD energy-momentum tensor (EMT). We propose different variants of GIRS, e.g., using specific values for the position vectors of the operators under study, or integrating over timeslices (t-GIRS), which may lead to reduced statistical noise in the nonperturbative calculations via lattice simulations. We provide results, up to one loop, for the conversion factors between the different versions of GIRS and the $\MSbar$ scheme.   

As future plans, GIRS and our proposed variants (t-GIRS, etc) could be immediately implemented on operators of similar kind, e.g., four-fermi operators and supersymmetric operators (Gluino-Glue, Noether supercurrent). 

\begin{acknowledgements}
M.C., H.P. and A.S. acknowledge financial support from the project \textit{``Quantum Fields on the Lattice''}, funded by the Cyprus Research and Innovation Foundation (RIF) under contract number EXCELLENCE/0918/0066. G.S. acknowledges financial support by the University of Cyprus, under the research programs entitled ``\textit{Quantum Fields on the Lattice}'' and ``\textit{Nucleon parton distribution functions using Lattice Quantum Chromodynamics}''. We thank C. Alexandrou, M. Dalla Brida, and K. Hadjiyiannakou for useful comments. 
\end{acknowledgements}

\appendix
\section{Technical aspects of the calculation}
\label{Technical aspects of the calculation}
There are three types of scalar Feynman integrals appearing in our calculation: 
\begin{eqnarray}
I_1 (\xi_1;\alpha_1) &\equiv & \ \int \frac{d^d p_1}{{(2 \pi)}^d} \frac{e^{i p_1 \cdot \xi_1}}{{(p^2)}^{\alpha_1}}, \\
I_2 (\xi_1, \xi_2;\alpha_1,\alpha_2,\alpha_3) &\equiv & \ \int \frac{d^d p_1 \ d^d p_2}{{(2 \pi)}^{2d}} \frac{e^{i p_1 \cdot \xi_1} \ e^{i p_2 \cdot \xi_2}}{{(p_1^2)}^{\alpha_1} \ {(p_2^2)}^{\alpha_2} \ {({(-p_1 + p_2)}^2)}^{\alpha_3}}, \\
I_3 (\xi_1, \xi_2;\alpha_1,\alpha_2,\alpha_3,\alpha_4,\alpha_5) &\equiv & \ \int \frac{d^d p_1 \ d^d p_2 \ d^d p_3}{{(2 \pi)}^{3d}} \frac{e^{i p_2 \cdot \xi_1} \ e^{i p_3 \cdot \xi_2}}{{(p_1^2)}^{\alpha_1} \ {(p_2^2)}^{\alpha_2} \ {({(-p_1 + p_2)}^2)}^{\alpha_3} \ {(p_3^2)}^{\alpha_4} \ {({(-p_1 + p_3)}^2)}^{\alpha_5}}.
\end{eqnarray}
For simplicity, we write down only scalar integrals for each type; integrands containing additional factors of ${p_1}_\mu$, ${p_2}_\nu$ can be handled in a similar way, or by taking derivatives of the results with respect to ${\xi_1}_\mu$, ${\xi_2}_\nu$, respectively. Below, we briefly describe the procedure for calculating each type of integral:
\begin{enumerate}
\item \textbf{Integral $I_1$:} \\
\\
We introduce Schwinger parameters:
\begin{equation}
\frac{1}{{(p_1^2)}^{\alpha_1}} = \frac{1}{\Gamma (\alpha_1)} \int_0^\infty d\lambda \ \lambda^{\alpha_1 - 1} e^{-\lambda p_1^2}.
\end{equation} 
After integrating over $p_1$ and $\lambda$, we get:
\begin{equation}
I_1 (\xi_1;\alpha_1) = \frac{\Gamma(-\alpha_1 + d/2) \ {(\xi_1^2)}^{\alpha_1 - d/2}}{4^\alpha_1 \ \pi^{d/2} \ \Gamma (\alpha_1)}.
\end{equation}
\clearpage
\item \textbf{Integral $I_2$:} \\
\\
We introduce Schwinger parameters:
\begin{eqnarray}
\frac{1}{{(p_1^2)}^{\alpha_1} \ {(p_2^2)}^{\alpha_2} \ {({(-p_1 + p_2)}^2)}^{\alpha_3}} &=& \frac{1}{\Gamma (\alpha_1) \ \Gamma (\alpha_2) \ \Gamma (\alpha_3)} \times \nonumber \\
&&  \int_0^\infty d\lambda_1 \int_0^\infty d\lambda_2 \int_0^\infty d\lambda_3 \ \lambda_1^{\alpha_1 - 1} \lambda_2^{\alpha_2 - 1} \lambda_3^{\alpha_3 - 1} e^{-\lambda_1 p_1^2 -\lambda_2 p_2^2 -\lambda_3 (-p_1 + p_2)^2}.
\end{eqnarray} 
After integrating over $p_1$ and $p_2$, we make a change of variables: $x_1 = \lambda_3 / (\lambda_2 + \lambda_3)$, $x_2 = 1 - \lambda_2 \lambda_3 / (\lambda_1 \lambda_2 + \lambda_1 \lambda_3 + \lambda_2 \lambda_3)$ and $x_3 = \lambda_1 \lambda_2 \lambda_3/ (\lambda_1 \lambda_2 + \lambda_1 \lambda_3 + \lambda_2 \lambda_3)$. Integrals over $x_2$ and $x_3$ can be calculated algebraically, while the remaining integration over $x_1$ cannot be obtained in a closed form (for general values of $\alpha_i$). The resulting expression takes the following form:
\begin{eqnarray}
I_2 (\xi_1, \xi_2;\alpha_1, \alpha_2, \alpha_3) &=& \frac{\Gamma(-\alpha_1 + d/2) \ \Gamma(-\alpha_1 - \alpha_2 - \alpha_3 + d)}{4^{\alpha_1 + \alpha_2 + \alpha_3} \ \pi^d \ \Gamma(\alpha_2) \ \Gamma(\alpha_3) \ \Gamma(d/2)} \ {(\xi_2^2)}^{\alpha_1 + \alpha_2 + \alpha_3 - d} \times \nonumber \\
&& \int_0^1 dx_1 \ \Bigg[{(1 - x_1)}^{-1 + \alpha_1 + \alpha_2 - d/2} \ x_1^{-1 + \alpha_1 + \alpha_3 - d/2} \times \nonumber \\
&& \qquad \qquad _2F_1 \left(-\alpha_1 + d/2, -\alpha_1 - \alpha_2 - \alpha_3 +d, d/2; -\frac{{(\xi_1 + x_1 \xi_2)}^2}{(1-x_1) \ x_1 \ \xi_2^2)}\right)\Bigg],
\label{I2}
\end{eqnarray}
where $(\alpha_1 + \alpha_2 + \alpha_3 - d) <0$, $(-\alpha_1 + d/2) >0$, $\alpha_1>0$. The next step is to examine whether the integration over $x_1$ and the limit of vanishing regulator ($\varepsilon \rightarrow 0$, $\varepsilon = 2 - d/2$) can be safely interchanged without leading to divergences. For this check, it is useful to express the hypergeometric function appearing in \eqref{I2} as a power series in $x_1$ and $(1-x_1)$, by applying an appropriate transformation formula (see~\cite{Gradshteyn:2007}). In case the interchange is indeed permissible, the integration over $x_1$ can be performed after taking the limit $\varepsilon \rightarrow 0$; in all other cases, it turns out that the hypergeometric function can be expressed in terms of simpler functions, allowing a direct integration over $x_1$. The Laurent expansion of the hypergeometric function $_2 F_1$ over $\varepsilon = 0$ has been performed with the help of the mathematica package ``HypExp'' introduced in Ref.~\cite{Huber:2005yg}.
\item \textbf{Integral $I_3$:} \\
\\
We introduce:
 \begin{equation}
 1 = \frac{1}{d} \sum_\rho \frac{\partial {p_1}_\rho}{\partial {p_1}_\rho}.
 \end{equation}
 After integrating by parts, we get the following recursive relation, which can eliminate inverse powers of $p_1^2$, or $p_2^2$, or $p_3^2$:
 \begin{eqnarray}
 I_3 (\xi_1, \xi_2; \alpha_1, \alpha_2, \alpha_3, \alpha_4, \alpha_5) &=& \frac{1}{-2 \alpha_1 - \alpha_3 -\alpha_5 + d} \cdot \nonumber \\
&& \hspace{-1cm} \Big[ \alpha_3 \Big( I_3(\xi_1, \xi_2; \alpha_1 - 1, \alpha_2, \alpha_3 + 1, \alpha_4, \alpha_5) - I_3(\xi_1, \xi_2; \alpha_1, \alpha_2 - 1, \alpha_3 + 1, \alpha_4, \alpha_5) \Big) + \nonumber \\
&& \hspace{-1cm} \ \ \alpha_5 \Big( I_3(\xi_1, \xi_2; \alpha_1 - 1, \alpha_2, \alpha_3, \alpha_4, \alpha_5 + 1) - I_3(\xi_1, \xi_2; \alpha_1, \alpha_2, \alpha_3, \alpha_4 - 1, \alpha_5 + 1) \Big) \Big]. \quad \quad
\label{I3value}
 \end{eqnarray}
In the case where $\alpha_1$, $\alpha_2$, $\alpha_4$ are positive integers, which is true in the computation at hand, an iterative implementation of Eq. \eqref{I3value} leads to terms with one propagator less. One momentum can then be integrated using a well-known one-loop formula (see Eqs. (A.1 -- A.2) in Ref.~\cite{Chetyrkin:1981qh}); the remaining integrals are of type 1 or 2.
\end{enumerate}

\bibliographystyle{elsarticle-num}    
\bibliography{GIRS_References}

\begin{thebibliography}{10}
\expandafter\ifx\csname url\endcsname\relax
  \def\url#1{\texttt{#1}}\fi
\expandafter\ifx\csname urlprefix\endcsname\relax\def\urlprefix{URL }\fi
\expandafter\ifx\csname href\endcsname\relax
  \def\href#1#2{#2} \def\path#1{#1}\fi

\bibitem{Bochicchio:1985xa}
M.~Bochicchio, L.~Maiani, G.~Martinelli, G.~C. Rossi, M.~Testa, {Chiral
  Symmetry on the Lattice with Wilson Fermions}, Nucl. Phys. B262 (1985) 331.
\newblock \href {https://doi.org/10.1016/0550-3213(85)90290-1}
  {\path{doi:10.1016/0550-3213(85)90290-1}}.

\bibitem{Martinelli:1994ty}
G.~Martinelli, C.~Pittori, C.~T. Sachrajda, M.~Testa, A.~Vladikas, {A General
  method for nonperturbative renormalization of lattice operators}, Nucl. Phys.
  B 445 (1995) 81--108.
\newblock \href {http://arxiv.org/abs/hep-lat/9411010}
  {\path{arXiv:hep-lat/9411010}}, \href
  {https://doi.org/10.1016/0550-3213(95)00126-D}
  {\path{doi:10.1016/0550-3213(95)00126-D}}.

\bibitem{Gimenez:2004me}
V.~Gimenez, L.~Giusti, S.~Guerriero, V.~Lubicz, G.~Martinelli, S.~Petrarca,
  J.~Reyes, B.~Taglienti, E.~Trevigne, {Non-perturbative renormalization of
  lattice operators in coordinate space}, Phys. Lett. B598 (2004) 227--236.
\newblock \href {http://arxiv.org/abs/hep-lat/0406019}
  {\path{arXiv:hep-lat/0406019}}, \href
  {https://doi.org/10.1016/j.physletb.2004.07.053}
  {\path{doi:10.1016/j.physletb.2004.07.053}}.

\bibitem{Jansen:1995ck}
K.~Jansen, C.~Liu, M.~Luscher, H.~Simma, S.~Sint, R.~Sommer, P.~Weisz,
  U.~Wolff, {Nonperturbative renormalization of lattice QCD at all scales},
  Phys. Lett. B 372 (1996) 275--282.
\newblock \href {http://arxiv.org/abs/hep-lat/9512009}
  {\path{arXiv:hep-lat/9512009}}, \href
  {https://doi.org/10.1016/0370-2693(96)00075-5}
  {\path{doi:10.1016/0370-2693(96)00075-5}}.

\bibitem{Dimopoulos:2018zef}
P.~Dimopoulos, G.~Herdo\'\i{}za, M.~Papinutto, C.~Pena, D.~Preti, A.~Vladikas,
  {Non-Perturbative Renormalisation and Running of BSM Four-Quark Operators in
  $N_f = 2$ QCD}, Eur. Phys. J. C 78~(7) (2018) 579.
\newblock \href {http://arxiv.org/abs/1801.09455} {\path{arXiv:1801.09455}},
  \href {https://doi.org/10.1140/epjc/s10052-018-6002-y}
  {\path{doi:10.1140/epjc/s10052-018-6002-y}}.

\bibitem{Joglekar:1975nu}
S.~D. Joglekar, B.~W. Lee, {General Theory of Renormalization of Gauge
  Invariant Operators}, Annals Phys. 97 (1976) 160.
\newblock \href {https://doi.org/10.1016/0003-4916(76)90225-6}
  {\path{doi:10.1016/0003-4916(76)90225-6}}.

\bibitem{Maas:2011se}
A.~Maas, {Describing gauge bosons at zero and finite temperature}, Other thesis
  (2013).
\newblock \href {http://arxiv.org/abs/1106.3942} {\path{arXiv:1106.3942}},
  \href {https://doi.org/10.1016/j.physrep.2012.11.002}
  {\path{doi:10.1016/j.physrep.2012.11.002}}.

\bibitem{Cucchieri:2018doy}
A.~Cucchieri, D.~Dudal, T.~Mendes, O.~Oliveira, M.~Roelfs, P.~J. Silva,
  {Faddeev-Popov Matrix in Linear Covariant Gauge: First Results}, Phys. Rev. D
  98~(9) (2018) 091504.
\newblock \href {http://arxiv.org/abs/1809.08224} {\path{arXiv:1809.08224}},
  \href {https://doi.org/10.1103/PhysRevD.98.091504}
  {\path{doi:10.1103/PhysRevD.98.091504}}.

\bibitem{Chetyrkin:1999pq}
K.~Chetyrkin, A.~Retey, {Renormalization and running of quark mass and field in
  the regularization invariant and MS-bar schemes at three loops and four
  loops}, Nucl. Phys. B 583 (2000) 3--34.
\newblock \href {http://arxiv.org/abs/hep-ph/9910332}
  {\path{arXiv:hep-ph/9910332}}, \href
  {https://doi.org/10.1016/S0550-3213(00)00331-X}
  {\path{doi:10.1016/S0550-3213(00)00331-X}}.

\bibitem{Ruijl:2017eht}
B.~Ruijl, T.~Ueda, J.~Vermaseren, A.~Vogt, {Four-loop QCD propagators and
  vertices with one vanishing external momentum}, JHEP 06 (2017) 040.
\newblock \href {http://arxiv.org/abs/1703.08532} {\path{arXiv:1703.08532}},
  \href {https://doi.org/10.1007/JHEP06(2017)040}
  {\path{doi:10.1007/JHEP06(2017)040}}.

\bibitem{Luthe:2017ttg}
T.~Luthe, A.~Maier, P.~Marquard, Y.~Schr{\"o}der, {The five-loop Beta function
  for a general gauge group and anomalous dimensions beyond Feynman gauge},
  JHEP 10 (2017) 166.
\newblock \href {http://arxiv.org/abs/1709.07718} {\path{arXiv:1709.07718}},
  \href {https://doi.org/10.1007/JHEP10(2017)166}
  {\path{doi:10.1007/JHEP10(2017)166}}.

\bibitem{Chetyrkin:2017bjc}
K.~Chetyrkin, G.~Falcioni, F.~Herzog, J.~Vermaseren, {Five-loop renormalisation
  of QCD in covariant gauges}, JHEP 10 (2017) 179, [Addendum: JHEP 12, 006
  (2017)].
\newblock \href {http://arxiv.org/abs/1709.08541} {\path{arXiv:1709.08541}},
  \href {https://doi.org/10.1007/JHEP10(2017)179}
  {\path{doi:10.1007/JHEP10(2017)179}}.

\bibitem{2018PhRvD..97h5016G}
J.~A. {Gracey}, R.~M. {Simms}, {Renormalization of QCD in the interpolating
  momentum subtraction scheme at three loops}, \prd 97~(8) (2018) 085016.
\newblock \href {http://arxiv.org/abs/1801.10415} {\path{arXiv:1801.10415}},
  \href {https://doi.org/10.1103/PhysRevD.97.085016}
  {\path{doi:10.1103/PhysRevD.97.085016}}.

\bibitem{2018arXiv181211818H}
F.~{Herzog}, S.~{Moch}, B.~{Ruijl}, T.~{Ueda}, J.~A.~M. {Vermaseren},
  A.~{Vogt}, {Five-loop contributions to low-N non-singlet anomalous dimensions
  in QCD}, arXiv e-prints (2018) arXiv:1812.11818\href
  {http://arxiv.org/abs/1812.11818} {\path{arXiv:1812.11818}}.

\bibitem{Baikov:2019zmy}
P.~Baikov, K.~Chetyrkin, {Transcendental structure of multiloop massless
  correlators and anomalous dimensions}, JHEP 10 (2019) 190.
\newblock \href {http://arxiv.org/abs/1908.03012} {\path{arXiv:1908.03012}},
  \href {https://doi.org/10.1007/JHEP10(2019)190}
  {\path{doi:10.1007/JHEP10(2019)190}}.

\bibitem{Gracey:2009da}
J.~A. Gracey, {Three loop anti-MS operator correlation functions for deep
  inelastic scattering in the chiral limit}, JHEP 04 (2009) 127.
\newblock \href {http://arxiv.org/abs/0903.4623} {\path{arXiv:0903.4623}},
  \href {https://doi.org/10.1088/1126-6708/2009/04/127}
  {\path{doi:10.1088/1126-6708/2009/04/127}}.

\bibitem{Chetyrkin:2010dx}
K.~G. Chetyrkin, A.~Maier, {Massless correlators of vector, scalar and tensor
  currents in position space at orders $\alpha_s^3$ and $\alpha_s^4$: Explicit
  analytical results}, Nucl. Phys. B844 (2011) 266--288.
\newblock \href {http://arxiv.org/abs/1010.1145} {\path{arXiv:1010.1145}},
  \href {https://doi.org/10.1016/j.nuclphysb.2010.11.007}
  {\path{doi:10.1016/j.nuclphysb.2010.11.007}}.

\bibitem{Cichy:2012is}
K.~Cichy, K.~Jansen, P.~Korcyl, {Non-perturbative renormalization in coordinate
  space for $N_f=2$ maximally twisted mass fermions with tree-level Symanzik
  improved gauge action}, Nucl. Phys. B865 (2012) 268--290.
\newblock \href {http://arxiv.org/abs/1207.0628} {\path{arXiv:1207.0628}},
  \href {https://doi.org/10.1016/j.nuclphysb.2012.08.006}
  {\path{doi:10.1016/j.nuclphysb.2012.08.006}}.

\bibitem{Tomii:2018zix}
M.~Tomii, N.~H. Christ, {$O(4)$-symmetric position-space renormalization of
  lattice operators}, Phys. Rev. D99~(1) (2019) 014515.
\newblock \href {http://arxiv.org/abs/1811.11238} {\path{arXiv:1811.11238}},
  \href {https://doi.org/10.1103/PhysRevD.99.014515}
  {\path{doi:10.1103/PhysRevD.99.014515}}.

\bibitem{Yang:2018bft}
Y.-B. Yang, M.~Gong, J.~Liang, H.-W. Lin, K.-F. Liu, D.~Pefkou, P.~Shanahan,
  {Nonperturbatively renormalized glue momentum fraction at the physical pion
  mass from lattice QCD}, Phys. Rev. D98~(7) (2018) 074506.
\newblock \href {http://arxiv.org/abs/1805.00531} {\path{arXiv:1805.00531}},
  \href {https://doi.org/10.1103/PhysRevD.98.074506}
  {\path{doi:10.1103/PhysRevD.98.074506}}.

\bibitem{Shanahan:2018pib}
P.~E. Shanahan, W.~Detmold, {Gluon gravitational form factors of the nucleon
  and the pion from lattice QCD}, Phys. Rev. D99~(1) (2019) 014511.
\newblock \href {http://arxiv.org/abs/1810.04626} {\path{arXiv:1810.04626}},
  \href {https://doi.org/10.1103/PhysRevD.99.014511}
  {\path{doi:10.1103/PhysRevD.99.014511}}.

\bibitem{Yang:2018nqn}
Y.-B. Yang, J.~Liang, Y.-J. Bi, Y.~Chen, T.~Draper, K.-F. Liu, Z.~Liu, {Proton
  Mass Decomposition from the QCD Energy Momentum Tensor}, Phys. Rev. Lett.
  121~(21) (2018) 212001.
\newblock \href {http://arxiv.org/abs/1808.08677} {\path{arXiv:1808.08677}},
  \href {https://doi.org/10.1103/PhysRevLett.121.212001}
  {\path{doi:10.1103/PhysRevLett.121.212001}}.

\bibitem{Alexandrou:2020sml}
C.~Alexandrou, S.~Bacchio, M.~Constantinou, J.~Finkenrath, K.~Hadjiyiannakou,
  K.~Jansen, G.~Koutsou, H.~Panagopoulos, G.~Spanoudes, {Complete flavor
  decomposition of the spin and momentum fraction of the proton using lattice
  QCD simulations at physical pion mass}, Phys. Rev. D101~(9) (2020) 094513.
\newblock \href {http://arxiv.org/abs/2003.08486} {\path{arXiv:2003.08486}},
  \href {https://doi.org/10.1103/PhysRevD.101.094513}
  {\path{doi:10.1103/PhysRevD.101.094513}}.

\bibitem{DallaBrida:2020gux}
M.~Dalla~Brida, L.~Giusti, M.~Pepe, {Non-perturbative definition of the QCD
  energy-momentum tensor on the lattice}, JHEP 04 (2020) 043.
\newblock \href {http://arxiv.org/abs/2002.06897} {\path{arXiv:2002.06897}},
  \href {https://doi.org/10.1007/JHEP04(2020)043}
  {\path{doi:10.1007/JHEP04(2020)043}}.

\bibitem{Ji:1998pc}
X.-D. Ji, {Off forward parton distributions}, J. Phys. G 24 (1998) 1181--1205.
\newblock \href {http://arxiv.org/abs/hep-ph/9807358}
  {\path{arXiv:hep-ph/9807358}}, \href
  {https://doi.org/10.1088/0954-3899/24/7/002}
  {\path{doi:10.1088/0954-3899/24/7/002}}.

\bibitem{Constantinou:2014tga}
M.~Constantinou, {Hadron Structure}, PoS LATTICE2014 (2015) 001.
\newblock \href {http://arxiv.org/abs/1411.0078} {\path{arXiv:1411.0078}},
  \href {https://doi.org/10.22323/1.214.0001} {\path{doi:10.22323/1.214.0001}}.

\bibitem{Larin:1993tq}
S.~A. Larin, {The Renormalization of the axial anomaly in dimensional
  regularization}, Phys. Lett. B303 (1993) 113--118.
\newblock \href {http://arxiv.org/abs/hep-ph/9302240}
  {\path{arXiv:hep-ph/9302240}}, \href
  {https://doi.org/10.1016/0370-2693(93)90053-K}
  {\path{doi:10.1016/0370-2693(93)90053-K}}.

\bibitem{Chanowitz:1979zu}
M.~S. Chanowitz, M.~Furman, I.~Hinchliffe, {The Axial Current in Dimensional
  Regularization}, Nucl. Phys. B159 (1979) 225--243.
\newblock \href {https://doi.org/10.1016/0550-3213(79)90333-X}
  {\path{doi:10.1016/0550-3213(79)90333-X}}.

\bibitem{tHooft:1972tcz}
G.~'t~Hooft, M.~J.~G. Veltman, {Regularization and Renormalization of Gauge
  Fields}, Nucl. Phys. B44 (1972) 189--213.
\newblock \href {https://doi.org/10.1016/0550-3213(72)90279-9}
  {\path{doi:10.1016/0550-3213(72)90279-9}}.

\bibitem{Siegel:1979wq}
W.~Siegel, {Supersymmetric Dimensional Regularization via Dimensional
  Reduction}, Phys. Lett. 84B (1979) 193--196.
\newblock \href {https://doi.org/10.1016/0370-2693(79)90282-X}
  {\path{doi:10.1016/0370-2693(79)90282-X}}.

\bibitem{Patel:1992vu}
A.~Patel, S.~R. Sharpe, {Perturbative corrections for staggered fermion
  bilinears}, Nucl. Phys. B395 (1993) 701--732.
\newblock \href {http://arxiv.org/abs/hep-lat/9210039}
  {\path{arXiv:hep-lat/9210039}}, \href
  {https://doi.org/10.1016/0550-3213(93)90054-S}
  {\path{doi:10.1016/0550-3213(93)90054-S}}.

\bibitem{Buras:1989xd}
A.~J. Buras, P.~H. Weisz, {QCD Nonleading Corrections to Weak Decays in
  Dimensional Regularization and 't Hooft-Veltman Schemes}, Nucl. Phys. B333
  (1990) 66--99.
\newblock \href {https://doi.org/10.1016/0550-3213(90)90223-Z}
  {\path{doi:10.1016/0550-3213(90)90223-Z}}.

\bibitem{Alles:1998is}
B.~All{\'e}s, A.~Feo, H.~Panagopoulos, {Asymptotic scaling corrections in QCD
  with Wilson fermions from the three loop average plaquette}, Phys. Lett. B
  426 (1998) 361--366, [Erratum: Phys.Lett.B 553, 337--338 (2003)].
\newblock \href {http://arxiv.org/abs/hep-lat/9801003}
  {\path{arXiv:hep-lat/9801003}}, \href
  {https://doi.org/10.1016/S0370-2693(98)00295-0}
  {\path{doi:10.1016/S0370-2693(98)00295-0}}.

\bibitem{Panagopoulos:2006ky}
H.~Panagopoulos, A.~Skouroupathis, A.~Tsapalis, {Free energy and plaquette
  expectation value for gluons on the lattice, in three dimensions}, Phys. Rev.
  D 73 (2006) 054511.
\newblock \href {http://arxiv.org/abs/hep-lat/0601009}
  {\path{arXiv:hep-lat/0601009}}, \href
  {https://doi.org/10.1103/PhysRevD.73.054511}
  {\path{doi:10.1103/PhysRevD.73.054511}}.

\bibitem{Athenodorou:2007hi}
A.~Athenodorou, H.~Panagopoulos, A.~Tsapalis, {The Lattice Free Energy of QCD
  with Clover Fermions, up to Three-Loops}, Phys. Lett. B 659 (2008) 252--259.
\newblock \href {http://arxiv.org/abs/0710.3856} {\path{arXiv:0710.3856}},
  \href {https://doi.org/10.1016/j.physletb.2007.11.064}
  {\path{doi:10.1016/j.physletb.2007.11.064}}.

\bibitem{Brambilla:2013sua}
M.~Brambilla, F.~Di~Renzo, {High-loop perturbative renormalization constants
  for Lattice QCD (II): three-loop quark currents for tree-level Symanzik
  improved gauge action and $n_f$=2 Wilson fermions}, Eur. Phys. J. C 73~(12)
  (2013) 2666.
\newblock \href {http://arxiv.org/abs/1310.4981} {\path{arXiv:1310.4981}},
  \href {https://doi.org/10.1140/epjc/s10052-013-2666-5}
  {\path{doi:10.1140/epjc/s10052-013-2666-5}}.

\bibitem{Constantinou:2017sej}
M.~Constantinou, H.~Panagopoulos, {Perturbative renormalization of quasi-parton
  distribution functions}, Phys. Rev. D96~(5) (2017) 054506.
\newblock \href {http://arxiv.org/abs/1705.11193} {\path{arXiv:1705.11193}},
  \href {https://doi.org/10.1103/PhysRevD.96.054506}
  {\path{doi:10.1103/PhysRevD.96.054506}}.

\bibitem{Gracey:2003yr}
J.~A. Gracey, {Three loop anomalous dimension of nonsinglet quark currents in
  the RI-prime scheme}, Nucl. Phys. B662 (2003) 247--278.
\newblock \href {http://arxiv.org/abs/hep-ph/0304113}
  {\path{arXiv:hep-ph/0304113}}, \href
  {https://doi.org/10.1016/S0550-3213(03)00335-3}
  {\path{doi:10.1016/S0550-3213(03)00335-3}}.

\bibitem{Freedman:1974gs}
D.~Z. Freedman, I.~J. Muzinich, E.~J. Weinberg, {On the Energy-Momentum Tensor
  in Gauge Field Theories}, Annals Phys. 87 (1974) 95.
\newblock \href {https://doi.org/10.1016/0003-4916(74)90448-5}
  {\path{doi:10.1016/0003-4916(74)90448-5}}.

\bibitem{Ji:1995sv}
X.-D. Ji, {Breakup of hadron masses and energy - momentum tensor of QCD}, Phys.
  Rev. D 52 (1995) 271--281.
\newblock \href {http://arxiv.org/abs/hep-ph/9502213}
  {\path{arXiv:hep-ph/9502213}}, \href
  {https://doi.org/10.1103/PhysRevD.52.271}
  {\path{doi:10.1103/PhysRevD.52.271}}.

\bibitem{Ji:1996ek}
X.-D. Ji, {Gauge-Invariant Decomposition of Nucleon Spin}, Phys. Rev. Lett. 78
  (1997) 610--613.
\newblock \href {http://arxiv.org/abs/hep-ph/9603249}
  {\path{arXiv:hep-ph/9603249}}, \href
  {https://doi.org/10.1103/PhysRevLett.78.610}
  {\path{doi:10.1103/PhysRevLett.78.610}}.

\bibitem{Radyushkin:1997ki}
A.~Radyushkin, {Nonforward parton distributions}, Phys. Rev. D 56 (1997)
  5524--5557.
\newblock \href {http://arxiv.org/abs/hep-ph/9704207}
  {\path{arXiv:hep-ph/9704207}}, \href
  {https://doi.org/10.1103/PhysRevD.56.5524}
  {\path{doi:10.1103/PhysRevD.56.5524}}.

\bibitem{Horsley:2012pz}
R.~Horsley, R.~Millo, Y.~Nakamura, H.~Perlt, D.~Pleiter, P.~Rakow,
  G.~Schierholz, A.~Schiller, F.~Winter, J.~Zanotti, {A Lattice Study of the
  Glue in the Nucleon}, Phys. Lett. B 714 (2012) 312--316.
\newblock \href {http://arxiv.org/abs/1205.6410} {\path{arXiv:1205.6410}},
  \href {https://doi.org/10.1016/j.physletb.2012.07.004}
  {\path{doi:10.1016/j.physletb.2012.07.004}}.

\bibitem{Aoyama:2020ynm}
T.~Aoyama, et~al., {The anomalous magnetic moment of the muon in the Standard
  Model} (6 2020).
\newblock \href {http://arxiv.org/abs/2006.04822} {\path{arXiv:2006.04822}}.

\bibitem{Caracciolo:1991cp}
S.~Caracciolo, P.~Menotti, A.~Pelissetto, {One loop analytic computation of the
  energy momentum tensor for lattice gauge theories}, Nucl. Phys. B 375 (1992)
  195--239.
\newblock \href {https://doi.org/10.1016/0550-3213(92)90339-D}
  {\path{doi:10.1016/0550-3213(92)90339-D}}.

\bibitem{Panagopoulos:2020qcn}
G.~Panagopoulos, H.~Panagopoulos, G.~Spanoudes, {Two-loop renormalization and
  mixing of gluon and quark energy-momentum tensor operators}, Phys. Rev. D 103
  (2021) 014515.
\newblock \href {http://arxiv.org/abs/2010.02062} {\path{arXiv:2010.02062}},
  \href {https://doi.org/10.1103/PhysRevD.103.014515}
  {\path{doi:10.1103/PhysRevD.103.014515}}.

\bibitem{Gradshteyn:2007}
I.~Gradshteyn, I.~M. Ryzhik, {Tables of integrals, series, and products},
  Elsevier, 2007, Equation 9.132.1.

\bibitem{Huber:2005yg}
T.~Huber, D.~Maitre, {HypExp: A Mathematica package for expanding
  hypergeometric functions around integer-valued parameters}, Comput. Phys.
  Commun. 175 (2006) 122--144.
\newblock \href {http://arxiv.org/abs/hep-ph/0507094}
  {\path{arXiv:hep-ph/0507094}}, \href
  {https://doi.org/10.1016/j.cpc.2006.01.007}
  {\path{doi:10.1016/j.cpc.2006.01.007}}.

\bibitem{Chetyrkin:1981qh}
K.~Chetyrkin, F.~Tkachov, {Integration by Parts: The Algorithm to Calculate
  beta Functions in 4 Loops}, Nucl. Phys. B 192 (1981) 159--204.
\newblock \href {https://doi.org/10.1016/0550-3213(81)90199-1}
  {\path{doi:10.1016/0550-3213(81)90199-1}}.

\end{thebibliography}

\end{document}